# Quantum chemistry based on classical mechanics inspired by simulated bifurcation

Fumihiko Aiga[1]* and Hayato Goto[1,2]

[1]Corporate Laboratory, Toshiba Corporation, Kawasaki, Kanagawa 212-8582, Japan
[2]RIKEN Center for Quantum Computing (RQC), Wako, Saitama 351-0198, Japan
*Correspondence to: fumihiko.aiga.x56@mail.toshiba

## Abstract

Accurate quantum chemical calculations are critical for understanding molecular properties, yet their computational cost remains a major challenge. Full Configuration Interaction (FCI) provides exact solutions but is prohibitively expensive for large systems. To address this, quantum computers are expected to be useful, but developing practical quantum computers is still ongoing. Here we introduce an efficient Configuration Interaction (CI) computation algorithm based on classical mechanics, which we call Simulated Bifurcation-based CI (SBCI), because we derive this algorithm from a quantum inspired algorithm for combinatorial optimization called Simulated Bifurcation. Applying it to FCI computations of representative molecular systems and comparing the results with those by a standard method, we demonstrate that SBCI can reduce computation costs such as computation times and/or required memory sizes, while keeping high accuracy comparable to the standard method. Thus, SBCI will be promising for accelerating high-precision electronic structure calculations without compromising reliability.

## Introduction

In quantum chemical calculations, we seek the solution of the time-independent Schrödinger equation for the electronic states of molecules in the Born-Oppenheimer approximation. For describing systems with strong electron correlation, such as transition metal complexes or bond dissociations, configuration interaction (CI) (*1*) is essential. In full CI (FCI), which is exact within a basis set, the number of configurations increases combinatorially in the number of electrons and orbitals. Large CI problems on classical computers have been explored (*2−4*), and their implementation on graphic processing units has also been developed (*5*). For system sizes beyond the reach of FCI, we have to employ an approximate ansatz, such as truncated CI (*1*), complete active space self-consistent field (*6*), and those plus perturbative corrections (*7−9*). On the other hand, the advantage of gate-based quantum computers for quantum chemical calculations is anticipated, and various algorithms have been proposed (*10,11*). In the hybrid quantum-classical algorithms called quantum-selected CI (*12*) or sample-based quantum diagonalization (*13*), important configurations are identified by sampling on quantum computers, and the Hamiltonian in the subspace spanned by those important configurations is diagonalized on classical computers. However, practical-size reliable quantum computers are still under development.

In the above CI calculations on classical computers (*2–9,12,13*), the Davidson method (*14,15*) is commonly used for eigenvalue computations. Efforts have been made to improve the preconditioning in the Davidson method (*16*) and to reduce the number of stored vectors (*17,18*). Furthermore, alternative methods such as locally optimal block preconditioned conjugate gradient (LOBPCG) (*19,20*) have also been explored. Developing algorithms on classical computers with lower computation costs (computation times and/or required memory sizes) than and the same level of accuracy as the Davidson method is crucial to enhance the practicability of CI calculations in scientific and technological applications.

Here, we propose a new algorithm, which we call simulated bifurcation-based CI (SBCI), for efficiently solving CI eigenvalue problems of both ground and excited states on classical computers. Simulated Bifurcation (SB) algorithms have originally been proposed and developed for accelerating combinatorial optimization such as Ising problems (*21*) by one of the authors and coworkers (*22–28*). Inspired by SB, in the SBCI framework, we solve the optimization problem of the CI coefficients by mapping it onto classical mechanical dynamics. In SBCI, CI coefficients are regarded as the positions of the classical system, and their conjugate momenta are introduced to construct a classical Hamiltonian. As in SB, Hamilton's equations of motion are solved by the symplectic Euler method (*29*).

SBCI is fundamentally different from approaches based on Ising machines (*21, 30–32*) where the eigenvalue problem is transformed into quadratic unconstrained binary optimization or the Ising problem. Also, SBCI targets exact diagonalization and thus differs in direction from methods like the density matrix renormalization group (*33*) and FCI quantum Monte Carlo (*34*).

Interestingly, SBCI is similar to the method proposed by Car and Parrinello (*35*) in the sense that it maps the first-principles molecular dynamics simulations onto classical mechanical systems. This approach has been applied not only to periodic boundary systems analyzed by the Kohn-Sham equation but also to molecular systems analyzed by the wavefunction-based method (*36–38*). The Car-Parrinello method is formulated within a Lagrangian framework, wherein the time evolution of the system is executed via numerical integration schemes such as the Verlet algorithm. The electronic state based on the Kohn-Sham equation for periodic boundary systems may be solved by the conjugate gradient method (*39–41*). On the other hand, the trial wave function in SBCI evolves according to Hamilton's equations of motion using the symplectic Euler method. In addition, a distinctive feature of SBCI is that its trial wave function is not normalized, allowing for adaptive restarts triggered by the norm of the wave function, as explained later.

In usual CI approaches to determining excited states, it is standard to employ the Davidson method (*1,14,15*), wherein all target states are updated simultaneously. This method allows rapid convergence, particularly when calculating nearly degenerate states (*15*). However, when updating all target states concurrently, we must deal with the history of residual vectors for all unconverged states, resulting in increased computational costs for inner products and matrix-product constructions involving these vectors. Based on these considerations, we first introduce a single-state update algorithm, referred to as SBCI1. In SBCI1, by updating only one state, target states are determined sequentially from the lowest upward, leading to reduced computational costs. Second, we introduce a two-state update algorithm, referred to as SBCI2. In SBCI2, by updating two states simultaneously, target states are determined one by one from the lowest upward. Although SBCI2 requires higher costs than SBCI1, we expect that SBCI2 will offer the advantage of accelerated convergence over SBCI1, in particular, when dealing with nearly degenerate states. We implemented SBCI1 and SBCI2 into Python-based Simulations of Chemistry Framework (PySCF) (*42*), and performed several FCI computations. The energy values obtained by SBCI were in the same accuracy as those obtained by the Davidson method in PySCF, and both the values were consistent with those reported in the literature (*43–49*). In these examples, SBCI reduced computation time and memory usage compared to the Davidson method. Hence, SBCI has the potential to replace the Davidson method and become a new standard for future CI calculations.

**Results**

**SBCI1 algorithm**

Consider the electronic Hamiltonian matrix $H$ of a molecule, represented in terms of Slater determinants (*50*). The overall dimension is determined by the number of the present Slater determinants. When spatial symmetry is taken into account, we deal with the subspace corresponding to the irreducible representations of the point group. Figure 1A shows the SBCI1 procedure for obtaining the lowest $n$ eigenvalues and corresponding eigenvectors of $H$. To compute $n$ states from state 0 to state $n-1$, these states are determined sequentially from 0 upward. We compute one target state $\alpha$ at a time by updating only one state $\alpha$. Upon convergence for state $\alpha$, we obtain its eigenvalue $E_c^\alpha$ and eigenvector $x_c^\alpha$.

SBCI1 is as follows. To seek the $\alpha$-th eigenvector and eigenvalue of $H$, we first prepare an initial trial vector $x_0^\alpha$ (see sections S1.1 and S1.4 of the Supplementary Materials for details). Then, we introduce the *classical mechanical Hamiltonian* $H_{\text{SBCI1}}$ as

$$H_{\text{SBCI1}} = \frac{b_t^\alpha}{2} y'^{\alpha T}_t M^{-1} y'^\alpha_t + \frac{c_t^\alpha}{2} \frac{x_t^{\alpha T} H x_t^\alpha}{x_t^{\alpha T} x_t^\alpha} \tag{1}$$

where $x_t^\alpha$ and $y'^\alpha_t$ are the position and momentum vectors, respectively, of this classical system at time $t$, $b_t^\alpha$ and $c_t^\alpha$ are the time-dependent real parameters, $T$ denotes the transpose, and $M^{-1}$ is the diagonal inverse mass matrix. We regard $x_t^\alpha$ as a new trial vector for the $\alpha$-th eigenvector. Note that the second term in Eq. 1, namely, the potential energy, is proportional to the Rayleigh quotient defined as

$$E_t^\alpha = \frac{x_t^{\alpha T} H x_t^\alpha}{x_t^{\alpha T} x_t^\alpha} \tag{2}$$

to be minimized in the present case, as SB algorithms introduce the potential energy proportional to a cost function to be minimized in a combinatorial optimization problem (*22–28*). Hamilton's equations of motion for this classical system are given by (*29*)

$$\dot{x}_t^\alpha = \nabla_{y'^\alpha_t} H_{\text{SBCI1}} = b_t^\alpha M^{-1} y'^\alpha_t \tag{3}$$

$$\dot{y}'^\alpha_t = -\nabla_{x_t^\alpha} H_{\text{SBCI1}} = -c_t^\alpha z'^\alpha_t \tag{4}$$

where the dots denote time derivatives, $\mathbf{z'}_t^\alpha$ is the residual vector defined as

$$\mathbf{z'}_t^\alpha = \frac{(H - E_t^\alpha I)\mathbf{x}_t^\alpha}{\mathbf{x}_t^{\alpha T}\mathbf{x}_t^\alpha} \tag{5}$$

and $I$ is the unit matrix. Introducing $\mathbf{y}_t^\alpha$ and $\mathbf{z}_t^\alpha$ as

$$\mathbf{y}_t^\alpha = M^{-1}\mathbf{y'}_t^\alpha \tag{6}$$
$$\mathbf{z}_t^\alpha = M^{-1}\mathbf{z'}_t^\alpha \tag{7}$$

Eqs. 3 and 4 are rewritten as

$$\dot{\mathbf{x}}_t^\alpha = b_t^\alpha \mathbf{y}_t^\alpha \tag{8}$$
$$\dot{\mathbf{y}}_t^\alpha = -c_t^\alpha \mathbf{z}_t^\alpha \tag{9}$$

respectively. We solve Eqs. 8 and 9 by the symplectic Euler method (*29*), where time is discretized with a time step $\Delta_t = 1$. The updating rule is as follows:

$$\mathbf{y}_{t+1}^\alpha = \mathbf{y}_t^\alpha - c_t^\alpha \mathbf{z}_t^\alpha \tag{10}$$
$$\mathbf{x}_{t+1}^\alpha = \mathbf{x}_t^\alpha + b_t^\alpha \mathbf{y}_{t+1}^\alpha \tag{11}$$

In the case of excited state ($\alpha \geq 1$), we orthogonalize $\mathbf{z}_t^\alpha$ against the previously determined vectors $\mathbf{x}_c^i (i = 0, \ldots, \alpha - 1)$ (see Eq. S31 of the Supplementary Materials for details), and redefine it as $\mathbf{z}_t^\alpha$ before updating $\mathbf{y}_t^\alpha$ according to Eq. 10. Consequently, the updated vectors $\mathbf{y}_{t+1}^\alpha$ and $\mathbf{x}_{t+1}^\alpha = \mathbf{x}_t^\alpha + b_t^\alpha (\mathbf{y}_t^\alpha - c_t^\alpha \mathbf{z}_t^\alpha)$ are orthogonal to $\mathbf{x}_c^i (i = 0, \ldots, \alpha - 1)$. The inverse mass matrix $M^{-1}$ is set to the standard precondition matrix $(D - E^0 I)^{-1}$ (*1,14*), where $D$ is the diagonal of $H$, and $E^0$ is the newest updated Rayleigh quotient for the ground state (state 0).

Unlike the original SB algorithm (*22–28*), SBCI does not employ adiabatic time evolutions. Instead, we variationally determine the parameters $b_t^\alpha$ and $c_t^\alpha$. That is, we determine $b_t^\alpha$ and $c_t^\alpha$ such that $\mathbf{x}_{t+1}^\alpha = \mathbf{x}_t^\alpha + b_t^\alpha (\mathbf{y}_t^\alpha - c_t^\alpha \mathbf{z}_t^\alpha)$ is parallel to the lowest-eigenvalue eigenvector of $H$ in the subspace $\{\mathbf{x}_t^\alpha, \mathbf{y}_t^\alpha, \mathbf{z}_t^\alpha\}$, which can be done by diagonalizing the $3 \times 3$ matrix obtained by representing $H$ using the three orthonormal basis vectors derived from the Gram–Schmidt orthonormalization of $\mathbf{x}_t^\alpha$, $\mathbf{y}_t^\alpha$, and $\mathbf{z}_t^\alpha$. The obtained lowest eigenvalue is $E_{t+1}^\alpha$. The detail of updating is provided in sections S1.2 and S1.3 of the Supplementary Materials. If the convergence criterion for the state $\alpha$ (Eq. S61 of the Supplementary Materials for details) is satisfied, $\mathbf{x}_{t+1}^\alpha$ is normalized and stored as $\mathbf{x}_c^\alpha$, and $E_{t+1}^\alpha$ is stored as $E_c^\alpha$.

As found from Eq. 11, the updated vector $\mathbf{x}_{t+1}^\alpha$ is formed by adding an update term to the current vector $\mathbf{x}_t^\alpha$ and therefore not normalized. This characteristic feature enables SBCI1 to facilitate adaptive restarts through the evaluation of $|b_t^\alpha|$ and $|\mathbf{x}_{t+1}^\alpha|$, leading to faster convergence. The detail of the restart is provided in section S1.5 of the Supplementary Materials. The pseudocode for SBCI1 is given as algorithm 1 in the Materials and Methods section.

In the Davidson method (*1,14,15*), it is necessary to retain the history of the residual vectors for all unconverged states during the update process. In contrast, SBCI1 requires storing only two vectors $\mathbf{z}_t^\alpha$ and $\mathbf{y}_t^\alpha$ during the update process, leading to smaller memory usage and shorter computational time for inner products and matrix-product constructions involving these vectors.

**SBCI2 algorithm**

Here, we describe the overview of the SBCI2 algorithm shown in Fig. 1B (see section S2.1-2.5 of the Supplementary Materials for details). We consider the same molecular Hamiltonian matrix $H$ as in SBCI1. To compute $n$ states from state 0 to state $n-1$, in SBCI2, we compute a pair of states $(\alpha, \alpha+1)$ simultaneously, where $\alpha$ increases one by one from 0 to $n-2$. Upon convergence for state $\alpha$, we obtain its eigenvalue $E_c^\alpha$ and eigenvector $\boldsymbol{x}_c^\alpha$, then we proceed to computing the next pair of states. For the highest state ($\alpha = n-1$), the procedure is switched to SBCI1 and we compute the single state.

For the simultaneous computing of the states $\alpha$ and $\alpha+1$, we first prepare initial trial vectors $\boldsymbol{x}_0^\alpha$ and $\boldsymbol{x}_0^{\alpha+1}$ (see section S2.2 of the Supplementary Materials for details). We define the trial vectors at time $t$ as $\boldsymbol{x}_t^\alpha$ and $\boldsymbol{x}_t^{\alpha+1}$ for the $\alpha$-th and $(\alpha+1)$-th eigenvectors, respectively, and introduce the corresponding momentum vectors $\boldsymbol{y'}_t^\alpha$ and $\boldsymbol{y'}_t^{\alpha+1}$. We also define $\boldsymbol{y}_t^\alpha$ and $\boldsymbol{y}_t^{\alpha+1}$ as Eq. 6, $\boldsymbol{z'}_t^\alpha$ and $\boldsymbol{z'}_t^{\alpha+1}$ as Eq. 5, and $\boldsymbol{z}_t^\alpha$ and $\boldsymbol{z}_t^{\alpha+1}$ as Eq. 7. Introducing the *classical mechanical Hamiltonian* $H_{\text{SBCI2}}$ as

$$H_{\text{SBCI2}} = \sum_{i=\alpha}^{\alpha+1} \sum_{j=\alpha}^{\alpha+1} \frac{b_t^{i,j}}{2} \boldsymbol{y'}_t^{iT} M^{-1} \boldsymbol{y'}_t^j + \sum_{i=\alpha}^{\alpha+1} \frac{c_t^{i,i}}{2} \frac{\boldsymbol{x}_t^{iT} H \boldsymbol{x}_t^i}{\boldsymbol{x}_t^{iT} \boldsymbol{x}_t^i}$$
$$+ \sum_{i=\alpha}^{\alpha+1} \sum_{j=\alpha(i\neq j)}^{\alpha+1} \left( \frac{c_t^{i,j}}{2} \boldsymbol{x}_t^{iT} H \boldsymbol{x}_t^j + \frac{d_t^{i,j}}{2} \boldsymbol{x}_t^{iT} \boldsymbol{x}_t^j + \frac{a_t^{i,j}}{2} \boldsymbol{x}_t^{iT} M \boldsymbol{x}_t^j \right) \tag{12}$$

the updating rule based on the symplectic Euler method is as follows:

$$\boldsymbol{y}_{t+1} = \boldsymbol{y}_t - C_t \boldsymbol{z}_t + A_t \boldsymbol{x}_t \tag{13}$$
$$\boldsymbol{x}_{t+1} = \boldsymbol{x}_t + B_t \boldsymbol{y}_{t+1} \tag{14}$$

where

$$\boldsymbol{x}_t = \begin{pmatrix} \boldsymbol{x}_t^\alpha \\ \boldsymbol{x}_t^{\alpha+1} \end{pmatrix}, \quad \boldsymbol{y}_t = \begin{pmatrix} \boldsymbol{y}_t^\alpha \\ \boldsymbol{y}_t^{\alpha+1} \end{pmatrix}, \quad \boldsymbol{z}_t = \begin{pmatrix} \boldsymbol{z}_t^\alpha \\ \boldsymbol{z}_t^{\alpha+1} \end{pmatrix} \tag{15}$$

and we have introduced the following three real matrices:

$$A_t = \begin{pmatrix} 0 & a_t^{\alpha,\alpha+1} \\ a_t^{\alpha+1,\alpha} & 0 \end{pmatrix}, B_t = \begin{pmatrix} b_t^{\alpha,\alpha} & b_t^{\alpha,\alpha+1} \\ b_t^{\alpha+1,\alpha} & b_t^{\alpha+1,\alpha+1} \end{pmatrix}, C_t = \begin{pmatrix} c_t^{\alpha,\alpha} & c_t^{\alpha,\alpha+1} \\ c_t^{\alpha+1,\alpha} & c_t^{\alpha+1,\alpha+1} \end{pmatrix} \tag{16}$$

We orthogonalize $\boldsymbol{z}_t^\alpha$ and $\boldsymbol{z}_t^{\alpha+1}$ against the previously determined vectors $\boldsymbol{x}_c^i (i = 0, \ldots, \alpha-1)$ (see Eqs. S101a and S101b of the Supplementary Materials for details), and redefine them as $\boldsymbol{z}_t^\alpha$ and $\boldsymbol{z}_t^{\alpha+1}$ before updating $\boldsymbol{y}_t^\alpha$ and $\boldsymbol{y}_t^{\alpha+1}$ according to Eq. 13. Consequently, the updated vectors $\boldsymbol{y}_{t+1}^\alpha, \boldsymbol{y}_{t+1}^{\alpha+1}, \boldsymbol{x}_{t+1}^\alpha$, and $\boldsymbol{x}_{t+1}^{\alpha+1}$ are orthogonal to $\boldsymbol{x}_c^i (i = 0, \ldots, \alpha-1)$. As in SBCI1, we use $(D - E^0 I)^{-1}$ as the inverse mass matrix $M^{-1}$.

Similarly to SBCI1, we variationally determine the matrices $A_t$, $B_t$, and $C_t$ such that $\boldsymbol{x}_{t+1}^\alpha$ and $\boldsymbol{x}_{t+1}^{\alpha+1}$ are parallel to the lowest-eigenvalue and the next-lowest-eigenvalue eigenvectors, respectively, of $H$ in the subspace $\{\boldsymbol{x}_t^\alpha, \boldsymbol{x}_t^{\alpha+1}, \boldsymbol{y}_t^\alpha, \boldsymbol{y}_t^{\alpha+1}, \boldsymbol{z}_t^\alpha, \boldsymbol{z}_t^{\alpha+1}\}$, which can be done by diagonalizing the $6 \times 6$ matrix obtained by representing $H$ using the six orthonormal basis vectors derived from the canonical orthogonalization (*51*) of $\boldsymbol{x}_t^\alpha, \boldsymbol{x}_t^{\alpha+1}, \boldsymbol{y}_t^\alpha, \boldsymbol{y}_t^{\alpha+1}, \boldsymbol{z}_t^\alpha$, and $\boldsymbol{z}_t^{\alpha+1}$. The obtained lowest and second lowest eigenvalues are $E_{t+1}^\alpha$ and $E_{t+1}^{\alpha+1}$, respectively. If the convergence criterion for the state $\alpha$ (Eq. S127 of the Supplementary Materials for details) is satisfied, $\boldsymbol{x}_{t+1}^\alpha$ is normalized and stored as $\boldsymbol{x}_c^\alpha$, and $E_{t+1}^\alpha$ is stored as $E_c^\alpha$. The normalized $\boldsymbol{x}_{t+1}^{\alpha+1}$ and $E_{t+1}^{\alpha+1}$ are then used for the initial values for the lower state of the subsequent pair of states.

As well as in SBCI1, the trial CI wave functions $\boldsymbol{x}_t^\alpha$ and $\boldsymbol{x}_t^{\alpha+1}$ are not normalized, and therefore SBCI2 can facilitate adaptive restarts through the evaluation of $|b_t^{\alpha,\alpha}|$, $|b_t^{\alpha+1,\alpha+1}|$, $|\boldsymbol{x}_{t+1}^\alpha|$, and $|\boldsymbol{x}_{t+1}^{\alpha+1}|$, leading to faster convergence. The detail of restart is given in section S2.5 of the Supplementary Materials. The pseudocode for SBCI2 is given as algorithm 2 in the Materials and Methods section.

SBCI2 stores only four vectors $\boldsymbol{z}_t^\alpha$, $\boldsymbol{z}_t^{\alpha+1}$, $\boldsymbol{y}_t^\alpha$, and $\boldsymbol{y}_t^{\alpha+1}$ during the update process, leading to smaller memory usage and shorter computational time for inner products and matrix-product constructions than the Davidson method (*1,14,15*) for excited states. We expect that SBCI2 will offer the advantage of accelerated convergence over SBCI1 when dealing with nearly degenerate states.

**Performance of SBCI for FCI calculations**

First, we present the performance of SBCI1 for evaluating the ground states of $N_2$ and CN varying the distance between the two nuclei, namely, the bond length. Figure 2A shows the FCI potential energy curves of the ground states for $N_2$ by SBCI1 and the Davidson method in PySCF(*42*). The detailed setups for these instances are provided in section S.3.1 of the Supplementary Materials, and the details of the computer system are provided in the Materials and Methods section. Figure 2B shows the energies at the representative bond lengths in Fig 2A. Figure 2C shows the execution time of CI, $T_{CI}$ (sec), with the bond length in Fig. 2A. Figure 2D shows the values of $T_{CI}$ at the representative bond lengths in Fig. 2A. Figures 2E–2H show the similar results for CN. The values for all the bond lengths are provided in Tables S2 and S3 of the Supplementary Materials. The numbers of determinants, $N_{det}$, for $N_2$ and CN are $4.3\times10^9$ and $9.8\times10^8$, respectively. For both $N_2$ and CN, the execution times of CI for SBCI1 are shorter than those for the Davidson method at any bond length, and the performance advantage of SBCI1 increases as the bond length becomes larger. Figure 3A shows the residual energies from the convergence values at each step for $N_2$ at the largest bond length. Figures 3B and 3C show the parameters $b_t^\alpha$, $c_t^\alpha$, and the norm of the wavefunction $|x_t^\alpha|$ ($\alpha = 0$), respectively, in SBCI1 at each step $t$ for $N_2$. Figures 3D–3F show the similar results for CN. The adaptive restarts at the step where $|x_t^\alpha| > 1.2$ (Eq. S62b of the Supplementary Materials for details) occur twice in Fig. 3C and four times in Fig. 3F.

Next, we present the performance of SBCI1 and SBCI2 for the excited states of $H_2O$, HF, $N_2$, BH, and $C_2$. The detailed setups for these instances are provided in section S.3.1 of the Supplementary Materials. Figure 4A shows the execution time of CI, $T_{CI}$ (sec), and the amount of memory (GB) by SBCI1, SBCI2 and the Davidson method for each instance, which are also shown in Figs 4B and 4C, respectively. Figure 4B shows that SBCI1 and SBCI2 are faster than the Davidson method, and SBCI2 is the fastest for $H_2O$, HF, BH, and $C_2$. Furthermore, Fig. 4C shows that SBCI1 and SBCI2 enable lower memory usage than the Davidson method.

Then, we present the accuracy of SBCI1 and SBCI2. The energies for the ground states of $N_2$ and CN by SBCI1 are shown in Figs. 2B and 2F, respectively, and those for the excited states of $H_2O$, HF, $N_2$, BH, and $C_2$ by SBCI2 are shown in Table 1. From these results, we conclude that SBCI1 and SBCI2 can achieve the same accuracy as the Davidson method. Note that both the energies obtained by SBCI and the Davidson method are consistent with those reported in the literature (*43*, *44*, *46−48*).

To confirm the validity of SBCI, we also evaluated other instances. We evaluated the ground states of $H_2O^+$(*45*) and $F_2$ (*3*), and the excited states of Ne (*43*) and the same molecules as Fig. 4A for different irreducible representations. Across all the instances, SBCI exhibits lower computational costs (in terms of both execution time and memory consumption) than the Davidson method while keeping comparable accuracy. The details are provided in section S.3.1, Figs. S2–S9 and Tables S4–S19 of the Supplementary Materials.

**Discussion**

We have proposed efficient algorithms for CI calculations of both ground and excited states of molecules based on classical mechanics inspired by a quantum-inspired combinatorial optimization algorithm called simulated bifurcation (SB). We have introduced two algorithms: SBCI1 and SBCI2. In SBCI1, by updating only one state, target states are determined sequentially from the lowest upward. In SBCI2, by updating two states simultaneously, target states are determined one by one from the lowest upward, leading to faster convergence than SBCI1 when energies are nearly degenerate. We have implemented SBCI1 and SBCI2 into PySCF, and performed several FCI calculations. The energies obtained by SBCI1 and SBCI2 were in the same accuracy as those by the Davidson method in PySCF. For the potential energy curve of the ground state, the execution time of CI for SBCI1 was shorter than that for the Davidson method at any bond length, and that the performance advantage of SBCI1 increased as the bond length becomes larger. For excited states, SBCI2 was faster than the Davidson method for all the instances, and reduced memory usage.

In this study, SBCI was applied only to CI; however, SBCI can also be employed for optimizing CI coefficients in multiconfigurational self-consistent field calculations, including CASSCF (*2*,*42*). In this study, the standard preconditioner $(D - E^0 I)^{-1}$ was used as the inverse mass matrix $M^{-1}$; however, other preconditioners (*16*) can also be employed. In this study, the Hamiltonian matrix $H$ is represented in terms of Slater determinants; however, SBCI can also be applied to $H$ represented in terms of configuration state functions (*1*) which are the eigenfunctions of spin operator $\hat{S}^2$.

In SBCI, the parameters are determined variationally rather than through adiabatic time evolution, as in the original SB algorithm (*22−28*). This makes it difficult to investigate the properties of the underlying dynamical system. Nevertheless, we have obtained some insights into energy conservation in the dynamical system of SBCI1 (see section S4 of the Supplementary Materials for details).

An important future work is the multi-node implementation of SBCI with Message Passing Interface (MPI) to enable larger-scale, higher-speed computations. Since SBCI stores fewer vectors during the update process than the Davidson method, SBCI is expected to be advantageous for multi-node CI computations.

**Materials and Methods**

We implemented SBCI1 and SBCI2 into PySCF 2.7.0 (42) by rewriting the function named davidson1 in pyscf/lib/linalg_helper.py. All calculations were performed using AMD EPYC 7742 (64 core, 2.25 GHz) with 2048 GB of memory.

We examined the dependence of the CI computation time on the number of cores in the computer for Ne. Figure S1 of the Supplementary Materials shows the execution time of CI with the different number of cores for Ne ($D_{2h}$, Ag, $S_z$=0, $n$=9) by SBCI1, SBCI2, and the Davidson method. Numerical values are shown in Table S1 of the Supplementary Materials. SBCI2 achieves the shortest computation time for any number of cores, and for all methods, the computation time is minimized when the number of cores is 16. Based on these observations, all instances were computed using 16 cores for every method. We also used the standard preconditioner $(D - E^0 I)^{-1}$ for the Davidson method in PySCF.

---

**Algorithm 1: SBCI1 for solving state $\alpha$**

---

Input: Initial values $x_0^\alpha$, $E_0^\alpha$
Eigenvectors $\{x_c^i | i = 0, \ldots, \alpha - 1\}$ and eigenvalues $\{E_c^i | i = 0, \ldots, \alpha - 1\}$

1: $x_0^{\alpha\prime} = x_0^\alpha - \sum_{i=0}^{\alpha-1} x_c^i (x_c^{iT} \cdot x_0^\alpha)$, $\quad x_0^\alpha = \dfrac{x_0^{\alpha\prime}}{\sqrt{x_0^{\alpha\prime T} \cdot x_0^{\alpha\prime}}}$

2: $y_0^\alpha = \mathbf{0}, b_0^\alpha = 1$
3: $z_0^{\prime\alpha} = (H - E_0^\alpha I) x_0^\alpha$
4: $t = 0$
5: while $t < t_{max}$
6: $\quad z_t^\alpha = \left(1 - \sum_{i=0}^{\alpha-1} x_c^i x_c^{iT}\right)(D - E^0 I)^{-1} z_t^{\prime\alpha}$
7: $\quad$ mutually orthonormalize $x_t^\alpha, y_t^\alpha, z_t^\alpha$ by the Gram–Schmidt process
8: $\quad$ diagonalize $H$ in the subspace $\{x_t^\alpha, y_t^\alpha, z_t^\alpha\}$, and obtain $E_{t+1}^\alpha$ as the lowest eigenvalue
9: $\quad$ obtain $b_t^\alpha, c_t^\alpha$ using the lowest eigenvector in 8: (see Eqs. S24, S54, S55 of the Supplementary Materials for details)
10: $\quad y_{t+1}^\alpha = y_t^\alpha - c_t^\alpha z_t^\alpha$
11: $\quad x_{t+1}^\alpha = x_t^\alpha + b_t^\alpha y_{t+1}^\alpha$
12: $\quad z_{t+1}^{\prime\alpha} = \dfrac{1}{\sqrt{x_{t+1}^{\alpha T} \cdot x_{t+1}^\alpha}} (H - E_{t+1}^\alpha I) x_{t+1}^\alpha$
13: $\quad dE = |E_{t+1}^\alpha - E_t^\alpha|, \quad dz' = \sqrt{z_{t+1}^{\prime\alpha T} \cdot z_{t+1}^{\prime\alpha}}$
14: $\quad$ if the convergence criterion ($dE < \varepsilon_0$ and $dz' < r_0$) is satisfied
15: $\quad\quad x_c^\alpha = \dfrac{1}{\sqrt{x_{t+1}^{\alpha T} \cdot x_{t+1}^\alpha}} x_{t+1}^\alpha, \quad E_c^\alpha = E_{t+1}^\alpha$
16: $\quad\quad$ go to 23
17: $\quad$ else
18: $\quad\quad$ if the restart condition (any of Eq. S62a S62b or S62c of the Supplementary Materials) is true
19: $\quad\quad\quad x_0^\alpha = \dfrac{1}{\sqrt{x_{t+1}^{\alpha T} \cdot x_{t+1}^\alpha}} x_{t+1}^\alpha, \quad E_0^\alpha = E_{t+1}^\alpha$
20: $\quad\quad\quad$ go to 4
21: $\quad\quad$ end if

22:     end if
23: end while

---

Algorithm 2: SBCI2 for solving state $\alpha$

Input: Initial values $x_0^\alpha, E_0^\alpha, x_0^{\alpha+1}, E_0^{\alpha+1}$
      Eigenvectors $\{x_c^i | i = 0, ..., \alpha - 1\}$ and eigenvalues $\{E_c^i | i = 0, ..., \alpha - 1\}$

1: $x_0^{\alpha\prime} = x_0^\alpha - \sum_{i=0}^{\alpha-1} x_c^i(x_c^{iT} \cdot x_0^\alpha), \quad x_0^\alpha = \dfrac{x_0^{\alpha\prime}}{\sqrt{x_0^{\alpha\prime T} \cdot x_0^{\alpha\prime}}}$

$x_0^{\alpha+1\prime} = x_0^{\alpha+1} - \sum_{i=0}^{\alpha-1} x_c^i(x_c^{iT} \cdot x_0^{\alpha+1}), \quad x_0^{\alpha+1} = \dfrac{x_0^{\alpha+1\prime}}{\sqrt{x_0^{\alpha+1\prime T} \cdot x_0^{\alpha+1\prime}}}$

2: $y_0^\alpha = 0, \quad y_0^{\alpha+1} = 0, \quad B_0 = I$
3: $z_0^{\prime\alpha} = (H - E_0^\alpha I)x_0^\alpha, \quad z_0^{\prime\alpha+1} = (H - E_0^{\alpha+1}I)x_0^{\alpha+1}$
4: $t = 0$
5: while $t < t_{max}$
6:   $z_t^\alpha = \left(1 - \sum_{i=0}^{\alpha-1} x_c^i x_c^{iT}\right)(D - E^0 I)^{-1} z_t^{\prime\alpha}, \quad z_t^{\alpha+1} = \left(1 - \sum_{i=0}^{\alpha-1} x_c^i x_c^{iT}\right)(D - E^0 I)^{-1} z_t^{\prime\alpha+1}$
7:   mutually orthonormalize $x_t^\alpha, x_t^{\alpha+1}, y_t^\alpha, y_t^{\alpha+1}, z_t^\alpha, z_t^{\alpha+1}$ by canonical orthogonalization
8:   diagonalize $H$ in the subspace $\{x_t^\alpha, y_t^\alpha, z_t^\alpha, x_t^{\alpha+1}, y_t^{\alpha+1}, z_t^{\alpha+1}\}$,
     and obtain $E_{t+1}^\alpha$ and $E_{t+1}^{\alpha+1}$ as the lowest and the second lowest eigenvalues
9:   obtain $A_t, B_t, C_t$ using the lowest and the second lowest eigenvectors in 8
     (see Eqs. S112, S116, S117, S121 of the Supplementary Materials for details)
10:  $y_{t+1}^\alpha = y_t^\alpha - c_t^{\alpha,\alpha} z_t^\alpha - c_t^{\alpha,\alpha+1} z_t^{\alpha+1} + a_t^{\alpha,\alpha+1} x_t^{\alpha+1}$
     $y_{t+1}^{\alpha+1} = y_t^{\alpha+1} - c_t^{\alpha+1,\alpha} z_t^\alpha - c_t^{\alpha+1,\alpha+1} z_t^{\alpha+1} + a_t^{\alpha+1,\alpha} x_t^\alpha$
11:  $x_{t+1}^\alpha = x_t^\alpha + b_t^{\alpha,\alpha} y_{t+1}^\alpha + b_t^{\alpha,\alpha+1} y_{t+1}^{\alpha+1}$
     $x_{t+1}^{\alpha+1} = x_t^{\alpha+1} + b_t^{\alpha+1,\alpha} y_{t+1}^\alpha + b_t^{\alpha+1,\alpha+1} y_{t+1}^{\alpha+1}$
12:  $z_{t+1}^{\prime\alpha} = \dfrac{1}{\sqrt{x_{t+1}^{\alpha T} \cdot x_{t+1}^\alpha}}(H - E_{t+1}^\alpha I)x_{t+1}^\alpha, \quad z_{t+1}^{\prime\alpha+1} = \dfrac{1}{\sqrt{x_{t+1}^{\alpha+1 T} \cdot x_{t+1}^{\alpha+1}}}(H - E_{t+1}^{\alpha+1} I)x_{t+1}^{\alpha+1}$
13:  $dE = |E_{t+1}^\alpha - E_t^\alpha|, \quad dz' = \sqrt{z_{t+1}^{\prime\alpha T} \cdot z_{t+1}^{\prime\alpha}}$
14:  if the convergence criterion ($dE < \varepsilon_0$ and $dz' < r_0$) is satisfied
15:      $x_c^\alpha = \dfrac{1}{\sqrt{x_{t+1}^{\alpha T} \cdot x_{t+1}^\alpha}} x_{t+1}^\alpha, \quad E_c^\alpha = E_{t+1}^\alpha, \quad x_0^{\alpha+1} = \dfrac{1}{\sqrt{x_{t+1}^{\alpha+1 T} \cdot x_{t+1}^{\alpha+1}}} x_{t+1}^{\alpha+1}, \quad E_0^{\alpha+1} = E_{t+1}^{\alpha+1}$
16:      go to 23
17:  else
18:      if the restart condition (any of Eq. S130a S130b or S130c of the Supplementary
         Materials) is true
19:          $x_0^\alpha = \dfrac{1}{\sqrt{x_{t+1}^{\alpha T} \cdot x_{t+1}^\alpha}} x_{t+1}^\alpha, \quad E_0^\alpha = E_{t+1}^\alpha, \quad x_0^{\alpha+1} = \dfrac{1}{\sqrt{x_{t+1}^{\alpha+1 T} \cdot x_{t+1}^{\alpha+1}}} x_{t+1}^{\alpha+1}, \quad E_0^{\alpha+1} = E_{t+1}^{\alpha+1}$
20:          go to 4
21:      end if
22:  end if
23: end while

**Acknowledgments:** We thank Y. Kaneko, M. Watabiki, M. Iwasaki, K. Tatsumura, R. Hidaka, Y. Sakai, H. Chono, and K. Kubo for comments and support.

A

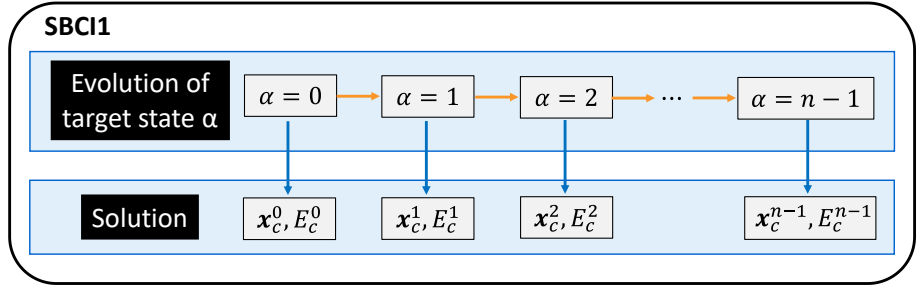

B

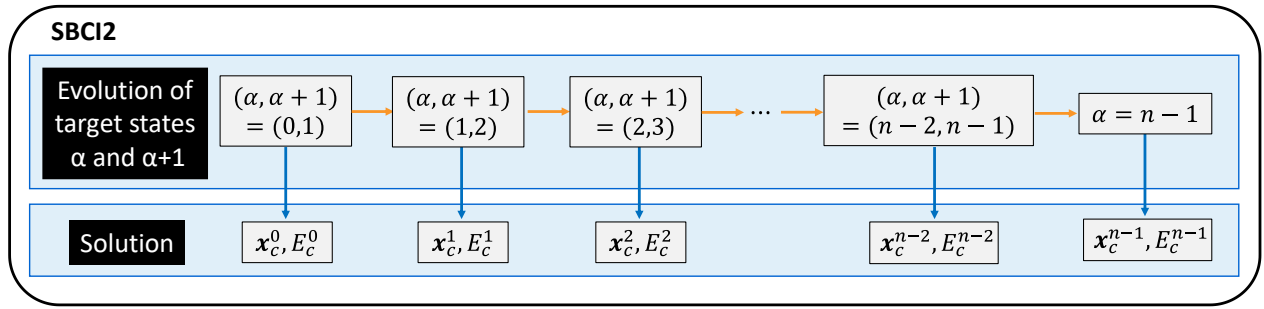

**Fig. 1. The procedure for obtaining the lowest $n$ states using SBCI1 and SBCI2.** (**A**) To compute $n$ states from state 0 to state $n-1$, in SBCI1, we compute one target state $\alpha$ at a time. The index $\alpha$ increases one by one from 0 to $n-1$. Upon convergence, we obtain its eigenvalue $E_c^\alpha$ and eigenvector $\boldsymbol{x}_c^\alpha$. (**B**) In SBCI2, we compute a pair of states $(\alpha, \alpha+1)$ simultaneously. The index $\alpha$ increases one by one from 0 to $n-2$. Upon convergence for state $\alpha$, we obtain its eigenvalue $E_c^\alpha$ and eigenvector $\boldsymbol{x}_c^\alpha$, then we proceed to computing the next pair of states. For the highest state ($\alpha = n-1$), the procedure switches to SBCI1 and we compute a single state.

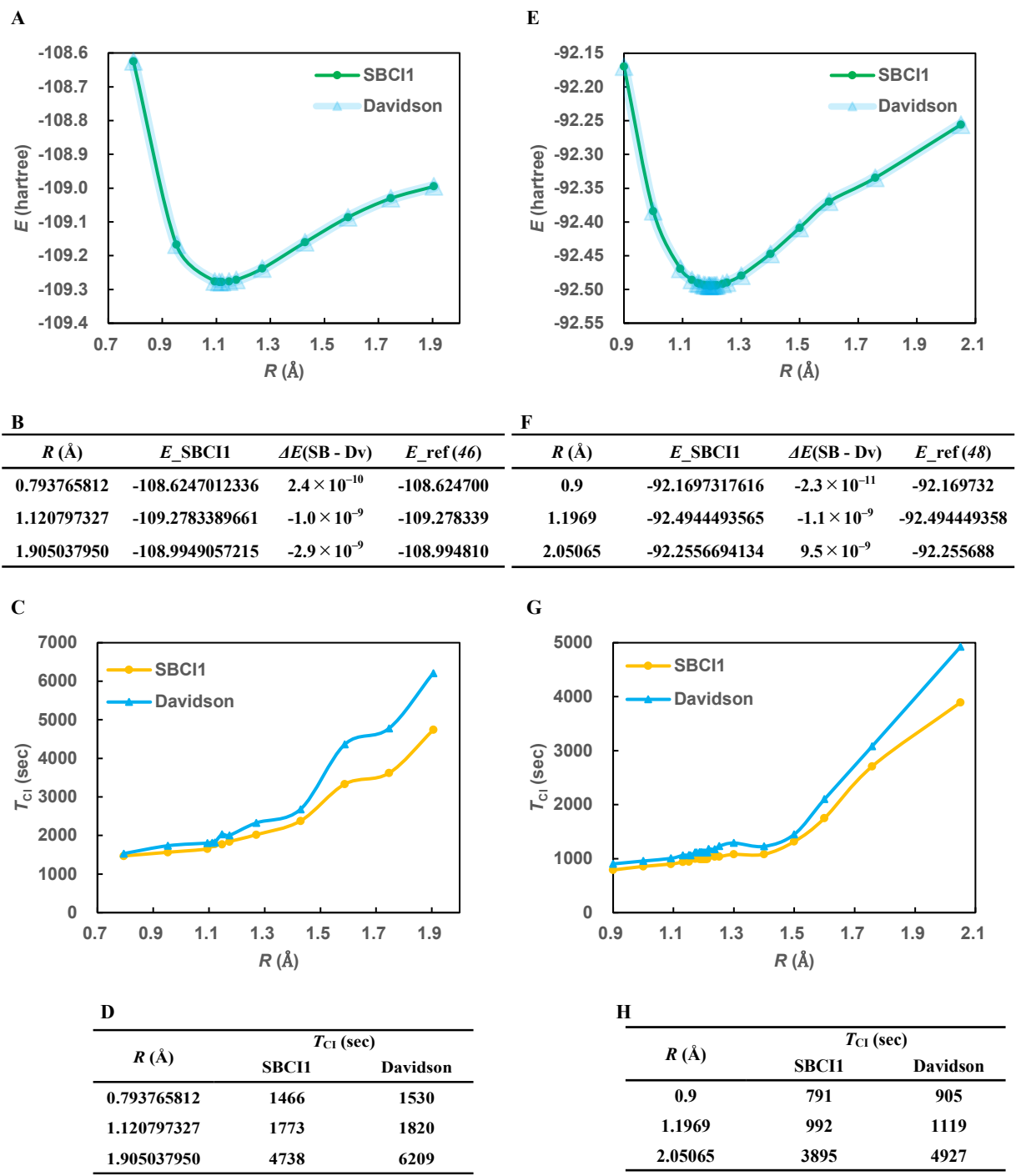

**Fig. 2. FCI potential energy curve of the ground state and execution time by SBCI1.** (**A**) FCI potential energy $E$ of the ground state for $N_2$ with the bond length $R$ by SBCI1 and the Davidson method. (**B**) FCI total energies $E\_SBCI1$ (hartree) at the representative bond lengths in (A). $\Delta E$(SB - Dv) (hartree) represents the difference obtained by subtracting the energy calculated using PySCF's Davidson method from $E\_SBCI1$. $E\_ref$ (hartree) is the energy in the literature (*46*). (**C**) The execution time of CI, $T_{CI}$, by SBCI1 and the Davidson method. (**D**) $T_{CI}$ in (C) at the representative bond lengths. (**E**−**H**) Similar results for CN.

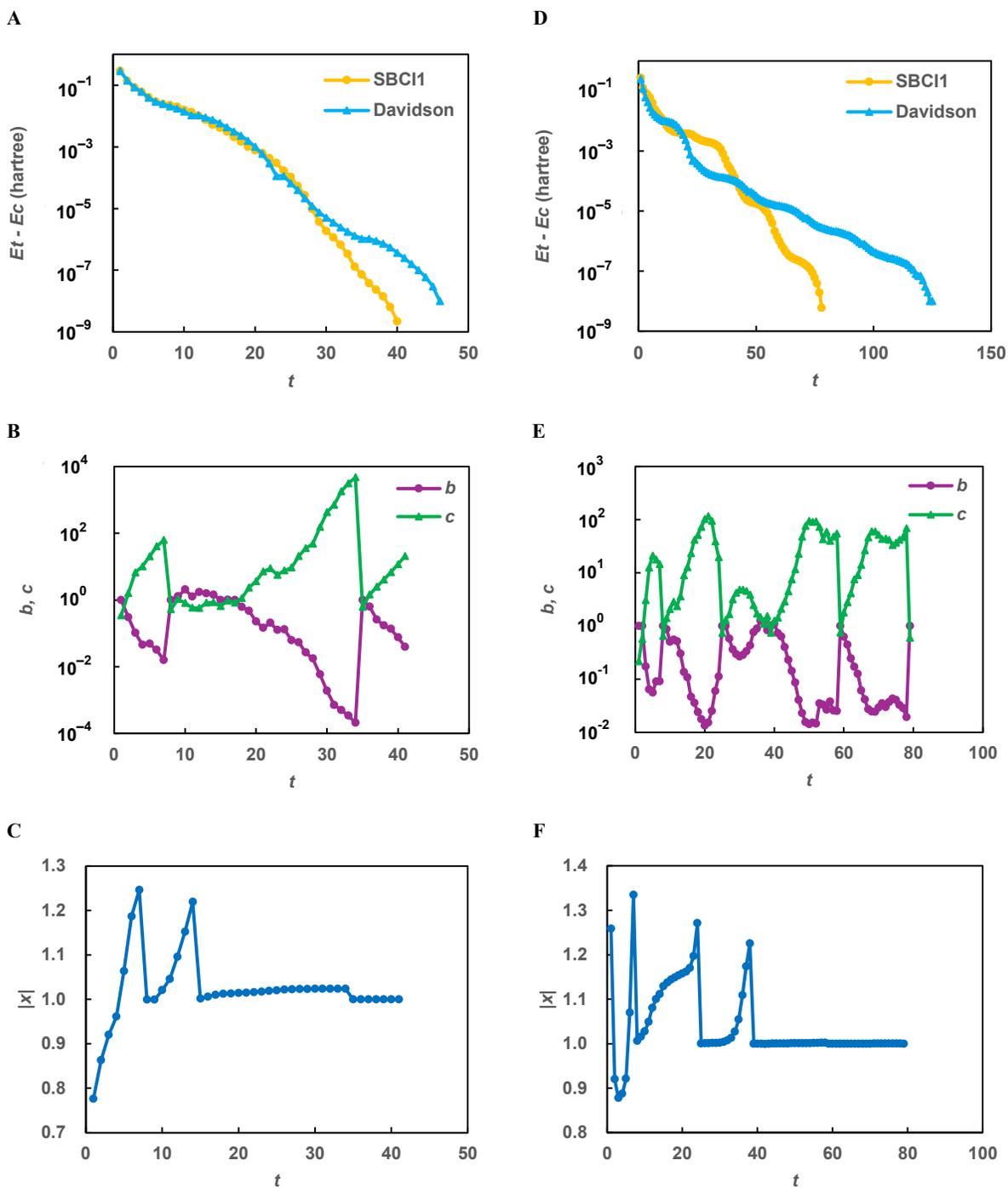

**Fig. 3. Time evolutions of FCI ground state energies and parameters in SBCI1.** (**A**) Residual energies, $E_t - E_c$, by SBCI1 and the Davidson method at each step $t$ for the ground state of $N_2$ at $R$= 1.905037950 Å. (**B**) Parameters $b_t^\alpha$, $c_t^\alpha$ in SBCI1 at each step $t$ in (A). (**C**) The norm of wavefunction $|x_t^\alpha|$ ($\alpha = 0$) in SBCI1 at each step $t$ in (A). (**D**–**F**) Similar results for CN at $R$= 2.05065 Å.

## A

| Instance | | $N_{det}$ | $T_{CI}$ (sec) | | | Memory (GB) | | |
|---|---|---|---|---|---|---|---|---|
| | | | SBCI1 | SBCI2 | Davidson | SBCI1 | SBCI2 | Davidson |
| $H_2O$ | $C_{2v}$, $A_1$, $S_z=0$, $n=4$ | $5.6 \times 10^8$ | 4683 | 4534 | 5336 | 38 | 48 | 73 |
| HF | $C_{2v}$, $A_2$, $S_z=1$, $n=3$ | $7.6 \times 10^8$ | 5109 | 4813 | 5571 | 46 | 56 | 77 |
| $N_2$ | $D_{\infty h}$, $E_{2uy}$, $S_z=0$, $n=4$ | $4.3 \times 10^9$ | 11355 | 14843 | 20375 | 212 | 246 | 609 |
| BH | $D_{\infty h}$, $E_{1x}$, $S_z=0$, $n=6$ | $2.0 \times 10^8$ | 14451 | 14068 | 17553 | 21 | 24 | 37 |
| $C_2$ | $D_{\infty h}$, $E_{2gx}$, $S_z=0$, $n=5$ | $2.2 \times 10^9$ | 16059 | 14077 | 28135 | 122 | 133 | 359 |

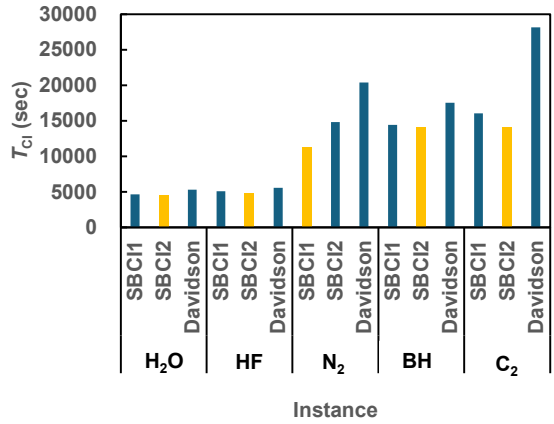
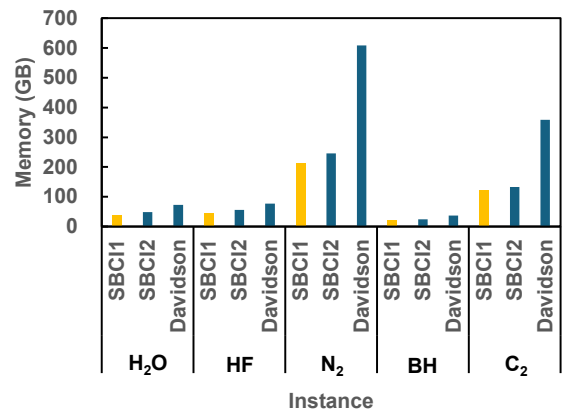

**Fig. 4. Performance of SBCI1 and SBCI2 for excited states.** (**A**) The execution time of CI, $T_{CI}$ (sec), and the amount of memory (GB) of SBCI1, SBCI2 and the Davidson method for each instance. $N_{det}$ is the number of determinants. (**B**) $T_{CI}$ (sec) for each instance. (**C**) The amount of memory (GB) for each instance.

**Table 1. Energies for the excited states obtained by SBCI2.** $S^2$ is the expectation value of the square of the spin operator. $E\_SBCI2$ (hartree) is FCI total energy obtained by SBCI2. $\Delta E$(SBCI2- Davidson) (hartree) represents the difference obtained by subtracting the energy calculated using PySCF's Davidson method from $E\_SBCI2$. $\Delta E$ex_SBCI2 and $\Delta E$ex_ref (eV for HF and hartree for others) are the excitation energies obtained by SBCI2 and those in the literature(*43*, *44*, *47*), respectively.

| \multicolumn{6}{l}{$H_2O$ $C_{2v}$, $A_1$, $S_z=0$} |
|---|---|---|---|---|---|
| State | $S^2$ | $E\_SBCI2$ | $\Delta E$(SBCI2 - Davidson) | $\Delta E_{ex}\_SBCI2$ | $\Delta E_{ex}\_ref$ (*44*) |
| 0 | 0 | -76.2582080336 | -2.9×10⁻¹¹ | | |
| 1 | 2 | -75.9112617436 | 4.0×10⁻⁹ | 0.346946 | 0.3470 |
| 2 | 0 | -75.8953399376 | 3.0×10⁻⁹ | 0.362868 | 0.3629 |
| 3 | 2 | -75.8603984356 | 7.9×10⁻¹⁰ | 0.397810 | 0.3978 |

| \multicolumn{6}{l}{HF $C_{2v}$, $A_2$, $S_z=1$} |
|---|---|---|---|---|---|
| State | $S^2$ | $E\_SBCI2$ | $\Delta E$(SBCI2 - Davidson) | $\Delta E_{ex}\_SBCI2$ (eV) | $\Delta E_{ex}\_ref$ (eV) (*47*) |
| 0 | 2 | -99.7155815109 | 4.6×10⁻¹⁰ | 14.926315 | 14.926 |
| 1 | 2 | -99.7037231002 | 8.5×10⁻¹⁰ | 15.248998 | 15.249 |

| \multicolumn{6}{l}{$N_2$ $D_{\infty h}$, $E_{2uy}$, $S_z=0$} |
|---|---|---|---|---|---|
| State | $S^2$ | $E\_SBCI2$ | $\Delta E$(SBCI2 - Davidson) | $\Delta E_{ex}\_SBCI2$ | $\Delta E_{ex}\_ref$ (*44,47*) |
| 0 | 2 | -108.9386467449 | 1.1×10⁻¹⁰ | 0.337881 | 0.3379 |
| 1 | 2 | -108.9090538486 | -6.4×10⁻¹⁰ | 0.367474 | 0.3675 |
| 2 | 0 | -108.8969563845 | -1.1×10⁻⁹ | 0.379571 | 0.3796 |
| 3 | 0 | -108.8826358743 | 3.2×10⁻¹⁰ | 0.393892 | 0.3939 |

| \multicolumn{6}{l}{BH $D_{\infty h}$, $E_{1x}$, $S_z=0$} |
|---|---|---|---|---|---|
| State | $S^2$ | $E\_SBCI2$ | $\Delta E$(SBCI2 - Davidson) | $\Delta E_{ex}\_SBCI2$ | $\Delta E_{ex}\_ref$ (*43*) |
| 0 | 2 | -25.1715009671 | 8.0×10⁻¹² | 0.048245 | |
| 1 | 0 | -25.1115701820 | 7.0×10⁻¹² | 0.108175 | 0.1082 |
| 2 | 2 | -24.9474437944 | 2.1×10⁻¹¹ | 0.272302 | |
| 3 | 0 | -24.9449962395 | 4.4×10⁻¹¹ | 0.274749 | 0.2744 |
| 4 | 2 | -24.9287051806 | 9.4×10⁻¹¹ | 0.291040 | |
| 5 | 0 | -24.9170020586 | -4.2×10⁻¹¹ | 0.302744 | 0.3028 |

| \multicolumn{6}{l}{$C_2$ $D_{\infty h}$, $E_{2gx}$, $S_z=0$} |
|---|---|---|---|---|---|
| State | $S^2$ | $E\_SBCI2$ | $\Delta E$(SBCI2 - Davidson) | $\Delta E_{ex}\_SBCI2$ | $\Delta E_{ex}\_ref$ (*44*) |
| 0 | 0 | -75.7302097784 | 1.5×10⁻⁹ | | |
| 1 | 0 | -75.6459483736 | 2.9×10⁻⁹ | 0.084261 | 0.08426 |

# Supplementary Text

## Section S1. Details of SBCI1

### S1.1 Initial trial vectors for SBCI1

Let us consider the task of determining the lowest $n$ eigenvalues and corresponding eigenvectors of the electronic Hamiltonian $H$ of a molecule. To construct the initial trial vectors, we select $n$ Slater determinants corresponding to the smallest $n$ diagonal elements of $H$, and prepare the corresponding vectors $\boldsymbol{x}_{-1}^i$ ($i = 0, \ldots, n-1$). These vectors are then subjected to Gram-Schmidt orthonormalization to verify their linear independence. If the norm of a vector, after subtracting its projection onto the already selected basis, falls below $1.0 \times 10^{-14}$, the corresponding Slater determinant is excluded. Then, we prepare

$$\boldsymbol{X}_{-1}^i = H \boldsymbol{x}_{-1}^i \tag{S1}$$

Using the eigenvectors $U_0$ obtained from the diagonalization of $H$ within the subspace $\{\boldsymbol{x}_{-1}^i\}$, we prepare

$$\boldsymbol{x}_0^\alpha = \sum_i U_0^{i,\alpha} \boldsymbol{x}_{-1}^i \tag{S2}$$

for initial trial vectors, and we prepare

$$\boldsymbol{X}_0^\alpha = H \boldsymbol{x}_0^\alpha = \sum_i U_0^{i,\alpha} \boldsymbol{X}_{-1}^i \tag{S3}$$

We also store the corresponding eigenvalues $\{E_0^i\}$. We use $\boldsymbol{x}_0^0$ ($\alpha = 0$) as the initial trial vector for the ground state. For the excited state ($\alpha \geq 1$), one may orthogonalize $\boldsymbol{x}_0^\alpha$ against the previously determined vectors $\boldsymbol{x}_c^i$ ($i = 0 \sim \alpha - 1$) and use it as the initial trial vector; however, in this work we employed the method described in Section S1.4 of the Supplementary Materials.

### S1.2 The first update step of SBCI1

Initially, we set $\boldsymbol{y}_0^\alpha = \boldsymbol{0}$. According to Eq. 11, the update formula is given by $\boldsymbol{x}_1^\alpha = \boldsymbol{x}_0^\alpha - b_0^\alpha c_0^\alpha \boldsymbol{z}_0^\alpha$, where we set $b_0^\alpha = 1$, and $c_0^\alpha$ is to be determined. If $\alpha$ corresponds to an excited state, then in order to orthogonalize $\boldsymbol{x}_0^\alpha$ against the previously obtained lower-state vectors $\boldsymbol{x}_c^i$ ($i = 0, \ldots, \alpha - 1$), we prepare

$$\boldsymbol{x}_0^{\alpha\prime} = \boldsymbol{x}_0^\alpha - \sum_{i=0}^{\alpha-1} \boldsymbol{x}_c^i (\boldsymbol{x}_c^{iT} \cdot \boldsymbol{x}_0^\alpha) \tag{S4}$$

$$\boldsymbol{X}_0^{\alpha\prime} = \boldsymbol{X}_0^\alpha - \sum_{i=0}^{\alpha-1} E_c^i \boldsymbol{x}_c^i (\boldsymbol{x}_c^{iT} \cdot \boldsymbol{x}_0^\alpha) \tag{S5}$$

where $E_c^i$ is the eigenvalue for the state $i$. Normalizing $\boldsymbol{x}_0^{\alpha\prime}$, we prepare

$$\boldsymbol{x}_0^\alpha = \frac{\boldsymbol{x}_0^{\alpha\prime}}{\sqrt{\boldsymbol{x}_0^{\alpha\prime T} \cdot \boldsymbol{x}_0^{\alpha\prime}}} \tag{S6}$$

$$\boldsymbol{X}_0^\alpha = \frac{\boldsymbol{X}_0^{\alpha\prime}}{\sqrt{\boldsymbol{x}_0^{\alpha\prime T} \cdot \boldsymbol{x}_0^{\alpha\prime}}} \tag{S7}$$

A preconditioned residual vector is constructed, and if $\alpha$ corresponds to an excited state, it is orthogonalized against the previously obtained lower-state vectors as follows:

$$\boldsymbol{z}_0^\alpha = \left(1 - \sum_{i=0}^{\alpha-1} \boldsymbol{x}_c^i \boldsymbol{x}_c^{iT}\right)(D - E^0 I)^{-1}(H - E_0^\alpha I)\boldsymbol{x}_0^\alpha \tag{S8}$$

We prepare

$$\boldsymbol{Z}_0^\alpha = H\boldsymbol{z}_0^\alpha \tag{S9}$$

Next, the vectors $\boldsymbol{x}_0^\alpha$ and $\boldsymbol{z}_0^\alpha$ are transformed into orthonormal basis vectors $\overline{\boldsymbol{x}}_0^\alpha$ and $\overline{\boldsymbol{z}}_0^\alpha$, respectively, using the Gram-Schmidt procedure as follows:

$$\overline{\boldsymbol{x}}_0^\alpha = \boldsymbol{x}_0^\alpha \tag{S10}$$

$$\overline{\boldsymbol{z}}_0^\alpha = \frac{\boldsymbol{z}_0^\alpha - (\overline{\boldsymbol{x}}_0^{\alpha T}\boldsymbol{z}_0^\alpha)\overline{\boldsymbol{x}}_0^\alpha}{|\boldsymbol{z}_0^\alpha - (\overline{\boldsymbol{x}}_0^{\alpha T}\boldsymbol{z}_0^\alpha)\overline{\boldsymbol{x}}_0^\alpha|} = B_z \boldsymbol{z}_0^\alpha + B_x \boldsymbol{x}_0^\alpha \tag{S11}$$

where

$$B_z = \frac{1}{\sqrt{N_{zz} - N_{xz}^2}} \tag{S12}$$

$$B_x = -N_{xz} B_z \tag{S13}$$

and the inner product of the vectors is expressed as $N_{zz} = \boldsymbol{z}_0^{\alpha T}\boldsymbol{z}_0^\alpha$. The $2 \times 2$ matrix $V_0$, representing the Hamiltonian $H$ in the subspace spanned by $\{\overline{\boldsymbol{x}}_0^\alpha, \overline{\boldsymbol{z}}_0^\alpha\}$, is given as follows:

$$V_0 = \begin{pmatrix} \overline{\boldsymbol{x}}_0^{\alpha T} H \overline{\boldsymbol{x}}_0^\alpha & \overline{\boldsymbol{x}}_0^{\alpha T} H \overline{\boldsymbol{z}}_0^\alpha \\ \overline{\boldsymbol{z}}_0^{\alpha T} H \overline{\boldsymbol{x}}_0^\alpha & \overline{\boldsymbol{z}}_0^{\alpha T} H \overline{\boldsymbol{z}}_0^\alpha \end{pmatrix} \tag{S14}$$

$$\overline{\boldsymbol{x}}_0^{\alpha T} H \overline{\boldsymbol{x}}_0^\alpha = H_{xx} \tag{S15a}$$

$$\overline{\boldsymbol{x}}_0^{\alpha T} H \overline{\boldsymbol{z}}_0^\alpha = \overline{\boldsymbol{z}}_0^{\alpha T} H \overline{\boldsymbol{x}}_0^\alpha = B_x H_{xx} + B_z H_{xz} \tag{S15b}$$

$$\overline{\boldsymbol{z}}_0^{\alpha T} H \overline{\boldsymbol{z}}_0^\alpha = B_x^2 H_{xx} + B_z^2 H_{zz} + 2B_x B_z H_{xz} \tag{S15c}$$

where the matrix elements are expressed as $H_{xx} = \boldsymbol{x}_0^{\alpha T} H \boldsymbol{x}_0^\alpha$. The lowest eigenvalue of the matrix $V_0$ is denoted as $E_1^\alpha$, and the corresponding eigenvector $\boldsymbol{v}$ is expressed as follows:

$$\boldsymbol{v} = v_x \overline{\boldsymbol{x}}_0^\alpha + v_z \overline{\boldsymbol{z}}_0^\alpha \tag{S16}$$

Substituting Eqs. S10 and S11 into Eq. S16, we obtain

$$\boldsymbol{v} = v_x \boldsymbol{x}_0^\alpha + v_z(B_z \boldsymbol{z}_0^\alpha + B_x \boldsymbol{x}_0^\alpha) = (v_x + v_z B_x)\boldsymbol{x}_0^\alpha + v_z B_z \boldsymbol{z}_0^\alpha \tag{S17}$$

On the other hand, according to Eqs. 10 and 11, the update formula is given as follows:

$$\boldsymbol{x}_1^\alpha = \boldsymbol{x}_0^\alpha - c_0^\alpha \boldsymbol{z}_0^\alpha \tag{S18}$$

The condition that $\boldsymbol{x}_1^\alpha$ should be parallel to $\boldsymbol{v}$ can be expressed as follows:

$$\boldsymbol{x}_1^\alpha = k_0^\alpha \boldsymbol{v} \tag{S19}$$

By substituting Eq. S18 into the left-hand side of Eq. S19, and substituting Eq. S17 into the right-hand side of Eq. S19, the following identity is obtained:

$$\boldsymbol{x}_0^\alpha - c_0^\alpha \boldsymbol{z}_0^\alpha = k_0^\alpha(v_x + v_z B_x)\boldsymbol{x}_0^\alpha + k_0^\alpha v_z B_z \boldsymbol{z}_0^\alpha \tag{S20}$$

From the equality of both sides of Eq. S20, the following is obtained:

$$1 = k_0^\alpha (v_x + v_z B_x) \tag{S21}$$
$$-c_0^\alpha = k_0^\alpha v_z B_z \tag{S22}$$

Solving Eq. S21 for $k_0^\alpha$ yields

$$k_0^\alpha = \frac{1}{v_x + v_z B_x} \tag{S23}$$

Substituting Eq. S23 into Eq. S22 yields

$$c_0^\alpha = -\frac{v_z B_z}{v_x + v_z B_x} \tag{S24}$$

Now, we can perform updating as follows:

$$\boldsymbol{y}_1^\alpha = -c_0^\alpha \boldsymbol{z}_0^\alpha \tag{S25}$$
$$\boldsymbol{Y}_1^\alpha = H\boldsymbol{y}_1^\alpha = -c_0^\alpha \boldsymbol{Z}_0^\alpha \tag{S26}$$
$$\boldsymbol{x}_1^\alpha = \boldsymbol{x}_0^\alpha + \boldsymbol{y}_1^\alpha \tag{S27}$$
$$\boldsymbol{X}_1^\alpha = H\boldsymbol{x}_1^\alpha = \boldsymbol{X}_0^\alpha + \boldsymbol{Y}_1^\alpha \tag{S28}$$

The residual vector is then constructed as follows:

$$\boldsymbol{z'}_1^\alpha = \frac{1}{k_0^\alpha}(H - E_1^\alpha I)\boldsymbol{x}_1^\alpha \tag{S29}$$

and if the convergence criteron

$$|E_1^\alpha - E_0^\alpha| < \varepsilon_0 \quad \text{and} \quad |\boldsymbol{z'}_1^\alpha| < r_0 \tag{S30}$$

is satisfied, then $\boldsymbol{x}_1^\alpha$ is normalized and stored as $\boldsymbol{x}_c^\alpha$. $E_1^\alpha$ is stored as $E_c^\alpha$, and the solution process for state $\alpha$ is terminated. If convergence is not achieved, the procedure continues as described in section S1.3. In this study, if frozen core=No, $\varepsilon_0 = 10^{-10}$ hartree and $r_0 = 10^{-5}$. If frozen core=Yes, $\varepsilon_0 = 10^{-8}$ hartree and $r_0 = 10^{-4}$.

## S1.3 The following update step of SBCI1

Consider time step $t(\geq 1)$. Assume that $\boldsymbol{x}_t^\alpha, \boldsymbol{X}_t^\alpha(= H\boldsymbol{x}_t^\alpha), \boldsymbol{y}_t^\alpha, \boldsymbol{Y}_t^\alpha(= H\boldsymbol{y}_t^\alpha)$, and $\boldsymbol{z'}_t^\alpha$ have been obtained. A preconditioned residual vector is constructed, and if $\alpha$ corresponds to an excited state, it is orthogonalized against the previously obtained lower-state vectors as follows:

$$\boldsymbol{z}_t^\alpha = \left(1 - \sum_{i=0}^{\alpha-1} \boldsymbol{x}_c^i \boldsymbol{x}_c^{iT}\right)(D - E^0 I)^{-1}\boldsymbol{z'}_t^\alpha \tag{S31}$$

We prepare

$$\boldsymbol{Z}_t^\alpha = H\boldsymbol{z}_t^\alpha \tag{S32}$$

Next, the vectors $\boldsymbol{x}_t^\alpha, \boldsymbol{y}_t^\alpha$, and $\boldsymbol{z}_t^\alpha$ are transformed into orthonormal basis vectors $\bar{\boldsymbol{x}}_t^\alpha, \bar{\boldsymbol{y}}_t^\alpha$, and $\bar{\boldsymbol{z}}_t^\alpha$, respectively, using the Gram-Schmidt procedure as follows:

$$\bar{\boldsymbol{x}}_t^\alpha = \frac{\boldsymbol{x}_t^\alpha}{|\boldsymbol{x}_t^\alpha|} = A_x \boldsymbol{x}_t^\alpha \tag{S33}$$

$$A_x = \frac{1}{|\boldsymbol{x}_t^\alpha|} = \frac{1}{\sqrt{N_{xx}}} \tag{S34}$$

$$\overline{\boldsymbol{y}}_t^\alpha = \frac{\boldsymbol{y}_t^\alpha - (\overline{\boldsymbol{x}}_t^{\alpha T}\boldsymbol{y}_t^\alpha)\overline{\boldsymbol{x}}_t^\alpha}{|\boldsymbol{y}_t^\alpha - (\overline{\boldsymbol{x}}_t^{\alpha T}\boldsymbol{y}_t^\alpha)\overline{\boldsymbol{x}}_t^\alpha|} = B_y \boldsymbol{y}_t^\alpha + B_x \boldsymbol{x}_t^\alpha \tag{S35}$$

$$B_y = \frac{1}{\sqrt{N_{yy} - \frac{N_{xy}^2}{N_{xx}}}} \tag{S36}$$

$$B_x = -\frac{N_{xy}}{N_{xx}} B_y \tag{S37}$$

$$\overline{\boldsymbol{z}}_t^\alpha = \frac{\boldsymbol{z}_t^\alpha - (\overline{\boldsymbol{x}}_t^{\alpha T}\boldsymbol{z}_t^\alpha)\overline{\boldsymbol{x}}_t^\alpha - (\overline{\boldsymbol{y}}_t^\alpha \boldsymbol{z}_t^\alpha)\overline{\boldsymbol{y}}_t^\alpha}{|\boldsymbol{z}_t^\alpha - (\boldsymbol{z}_t^\alpha)\overline{\boldsymbol{x}}_t^\alpha - (\overline{\boldsymbol{y}}_t^{\alpha T}\boldsymbol{z}_t^\alpha)\overline{\boldsymbol{y}}_t^\alpha|} = C_z \boldsymbol{z}_t^\alpha + C_x \boldsymbol{x}_t^\alpha + C_y \boldsymbol{y}_t^\alpha \tag{S38}$$

$$C_z = \frac{1}{\sqrt{N_{zz} - \frac{N_{xz}^2}{N_{xx}} - P^2}} \tag{S39}$$

$$C_y = -P B_y C_z \tag{S40}$$

$$C_x = -\left(\frac{N_{xz}}{N_{xx}} + P B_x\right) C_z \tag{S41}$$

$$P = B_x N_{xz} + B_y N_{zy} \tag{S42}$$

The $3 \times 3$ matrix $V_t$, representing the Hamiltonian $H$ in the subspace spanned by $\{\overline{\boldsymbol{x}}_t^\alpha, \overline{\boldsymbol{y}}_t^\alpha, \overline{\boldsymbol{z}}_t^\alpha\}$, is given as follows:

$$V_t = \begin{pmatrix} \overline{\boldsymbol{x}}_t^{\alpha T} H \overline{\boldsymbol{x}}_t^\alpha & \overline{\boldsymbol{x}}_t^{\alpha T} H \overline{\boldsymbol{y}}_t^\alpha & \overline{\boldsymbol{x}}_t^{\alpha T} H \overline{\boldsymbol{z}}_t^\alpha \\ \overline{\boldsymbol{y}}_t^{\alpha T} H \overline{\boldsymbol{x}}_t^\alpha & \overline{\boldsymbol{y}}_t^{\alpha T} H \overline{\boldsymbol{y}}_t^\alpha & \overline{\boldsymbol{y}}_t^{\alpha T} H \overline{\boldsymbol{z}}_t^\alpha \\ \overline{\boldsymbol{z}}_t^{\alpha T} H \overline{\boldsymbol{x}}_t^\alpha & \overline{\boldsymbol{z}}_t^{\alpha T} H \overline{\boldsymbol{y}}_t^\alpha & \overline{\boldsymbol{z}}_t^{\alpha T} H \overline{\boldsymbol{z}}_t^\alpha \end{pmatrix} \tag{S43}$$

$$\overline{\boldsymbol{x}}_t^{\alpha T} H \overline{\boldsymbol{x}}_t^\alpha = A_x^2 H_{xx} \tag{S44a}$$

$$\begin{aligned} \overline{\boldsymbol{x}}_t^{\alpha T} H \overline{\boldsymbol{y}}_t^\alpha &= \overline{\boldsymbol{y}}_t^{\alpha T} H \overline{\boldsymbol{x}}_t^\alpha \\ &= A_x \boldsymbol{x}_t^{\alpha T} H(B_y \boldsymbol{y}_t^\alpha + B_x \boldsymbol{x}_t^\alpha) \\ &= A_x (B_x H_{xx} + B_y H_{xy}) \end{aligned} \tag{S44b}$$

$$\begin{aligned} \overline{\boldsymbol{x}}_t^{\alpha T} H \overline{\boldsymbol{z}}_t^\alpha &= \overline{\boldsymbol{z}}_t^{\alpha T} H \overline{\boldsymbol{x}}_t^\alpha \\ &= A_x \boldsymbol{x}_t^{\alpha T} H(C_z \boldsymbol{z}_t^\alpha + C_x \boldsymbol{x}_t^\alpha + C_y \boldsymbol{y}_t^\alpha) \\ &= A_x (C_x H_{xx} + C_y H_{xy} + C_z H_{xz}) \end{aligned} \tag{S44c}$$

$$\begin{aligned} \overline{\boldsymbol{y}}_t^{\alpha T} H \overline{\boldsymbol{y}}_t^\alpha &= (B_y \boldsymbol{y}_t^{\alpha T} + B_x \boldsymbol{x}_t^{\alpha T}) H(B_y \boldsymbol{y}_t^\alpha + B_x \boldsymbol{x}_t^\alpha) \\ &= B_x^2 H_{xx} + B_y^2 H_{yy} + 2 B_x B_y H_{xy} \end{aligned} \tag{S44d}$$

$$\begin{aligned} \overline{\boldsymbol{y}}_t^{\alpha T} H \overline{\boldsymbol{y}}_t^\alpha &= \overline{\boldsymbol{z}}_t^{\alpha T} H \overline{\boldsymbol{y}}_t^\alpha \\ &= (B_y \boldsymbol{y}_t^{\alpha T} + B_x \boldsymbol{x}_t^{\alpha T}) H(C_z \boldsymbol{z}_t^\alpha + C_x \boldsymbol{x}_t^\alpha + C_y \boldsymbol{y}_t^\alpha) \\ &= B_x (C_x H_{xx} + C_y H_{xy} + C_z H_{xz}) + B_y (C_x H_{xy} + C_y H_{yy} + C_z H_{yz}) \end{aligned} \tag{S44e}$$

$$\bar{\mathbf{z}}_t^{\alpha T} H \bar{\mathbf{z}}_t^\alpha = (C_z \mathbf{z}_t^{\alpha T} + C_x \mathbf{x}_t^{\alpha T} + C_y \mathbf{y}_t^{\alpha T}) H (C_z \mathbf{z}_t^\alpha + C_x \mathbf{x}_t^\alpha + C_y \mathbf{y}_t^\alpha)$$
$$= C_x^2 H_{xx} + C_y^2 H_{yy} + C_z^2 H_{zz} + 2 C_x C_y H_{xy} + 2 C_x C_z H_{xz} + 2 C_y C_z H_{yz} \tag{S44f}$$

The lowest eigenvalue of the matrix $V_t$ is denoted as $E_{t+1}^\alpha$, and the corresponding eigenvector $\mathbf{v}$ is expressed as follows:

$$\mathbf{v} = v_x \bar{\mathbf{x}}_t^\alpha + v_y \bar{\mathbf{y}}_t^\alpha + v_z \bar{\mathbf{z}}_t^\alpha \tag{S45}$$

Substituting Eqs. S33, S35 and S38 into Eq. S45, we obtain

$$\mathbf{v} = v_x A_x \mathbf{x}_t^\alpha + v_y (B_y \mathbf{y}_t^\alpha + B_x \mathbf{x}_t^\alpha) + v_z (C_z \mathbf{z}_t^\alpha + C_x \mathbf{x}_t^\alpha + C_y \mathbf{y}_t^\alpha)$$
$$= (v_x A_x + v_y B_x + v_z C_x) \mathbf{x}_t^\alpha + (v_y B_y + v_z C_y) \mathbf{y}_t^\alpha + v_z C_z \mathbf{z}_t^\alpha \tag{S46}$$

On the other hand, substituting Eq. 10 into Eq. 11, the update formula is given as follows:
$$\mathbf{x}_{t+1}^\alpha = \mathbf{x}_t^\alpha + b_t^\alpha \mathbf{y}_t^\alpha - b_t^\alpha c_t^\alpha \mathbf{z}_t^\alpha \tag{S47}$$

The condition that $\mathbf{x}_{t+1}^\alpha$ should be parallel to $\mathbf{v}$ can be expressed as follows:
$$\mathbf{x}_{t+1}^\alpha = k_t^\alpha \mathbf{v} \tag{S48}$$

By substituting Eq. S47 into the left-hand side of Eq. S48, and substituting Eq. S46 into the right-hand side of Eq. S48, the following identity is obtained:

$$\mathbf{x}_t^\alpha + b_t^\alpha \mathbf{y}_t^\alpha - b_t^\alpha c_t^\alpha \mathbf{z}_t^\alpha = k_t^\alpha (v_x A_x + v_y B_x + v_z C_x) \mathbf{x}_t^\alpha + k_t^\alpha (v_y B_y + v_z C_y) \mathbf{y}_t^\alpha + k_t^\alpha v_z C_z \mathbf{z}_t^\alpha \tag{S49}$$

From the equality of both sides of Eq. S49, the following is obtained:
$$1 = k_t^\alpha (v_x A_x + v_y B_x + v_z C_x) \tag{S50}$$
$$b_t^\alpha = k_t^\alpha (v_y B_y + v_z C_y) \tag{S51}$$
$$-b_t^\alpha c_t^\alpha = k_t^\alpha v_z C_z \tag{S52}$$

Solving Eq. S50 for $k_t^\alpha$ yields

$$k_t^\alpha = \frac{1}{v_x A_x + v_y B_x + v_z C_x} \tag{S53}$$

Substituting Eq. S53 into Eq. S51 yields

$$b_t^\alpha = \frac{v_y B_y + v_z C_y}{v_x A_x + v_y B_x + v_z C_x} \tag{S54}$$

Solving Eq. S52 for $c_t^\alpha$, and using Eqs. S53 and S54, we obtain

$$c_t^\alpha = -\frac{v_z C_z}{v_y B_y + v_z C_y} \tag{S55}$$

Now, we can perform updating as follows:
$$\mathbf{y}_{t+1}^\alpha = \mathbf{y}_t^\alpha - c_t^\alpha \mathbf{z}_t^\alpha \tag{S56}$$
$$\mathbf{Y}_{t+1}^\alpha = H \mathbf{y}_{t+1}^\alpha = \mathbf{Y}_t^\alpha - c_t^\alpha \mathbf{Z}_t^\alpha \tag{S57}$$
$$\mathbf{x}_{t+1}^\alpha = \mathbf{x}_t^\alpha + b_t^\alpha \mathbf{y}_{t+1}^\alpha \tag{S58}$$
$$\mathbf{X}_{t+1}^\alpha = H \mathbf{x}_{t+1}^\alpha = \mathbf{X}_t^\alpha + b_t^\alpha \mathbf{Y}_{t+1}^\alpha \tag{S59}$$

The residual vector is then constructed as follows:

$$\mathbf{z'}_{t+1}^\alpha = \frac{1}{k_t^\alpha} (H - E_{t+1}^\alpha I) \mathbf{x}_{t+1}^\alpha \tag{S60}$$

and if the convergence criterion

$$|E_{t+1}^\alpha - E_t^\alpha| < \varepsilon_0 \quad \text{and} \quad |\mathbf{z'}_{t+1}^\alpha| < r_0 \tag{S61}$$

is satisfied, then $\mathbf{x}_{t+1}^\alpha$ is normalized and stored as $\mathbf{x}_c^\alpha$. $E_{t+1}^\alpha$ is stored as $E_c^\alpha$, and the solution process for state $\alpha$ is terminated. If convergence is not achieved, the procedure continues. In this study, if frozen core=No, $\varepsilon_0 = 10^{-10}$ hartree and $r_0 = 10^{-5}$. If frozen core=Yes, $\varepsilon_0 = 10^{-8}$ hartree and $r_0 = 10^{-4}$.

### S1.4 Initial trial vectors of SBCI1 for excited states

As the initial trial vector for the excited state $\alpha$, we employ the eigenvector corresponding to the second-lowest eigenvalue of matrix $V_0$ (Eq. S14) or matrix $V_t$ (Eq. S43), obtained when the state $\alpha - 1$ has converged. This vector is already orthogonal to the lower eigenvectors.

### S1.5 Restart of SBCI1

If convergence is not achieved under the criteria defined by Eq. S30 or Eq. S61a with Eq. 61b, each of the following conditions was evaluated:

$$\alpha > 0 \quad and \quad |b_t^\alpha| < b_{th} \quad and \quad |E_{t+1}^\alpha - E_t^\alpha| < \varepsilon_1 \tag{S62a}$$

$$|k_t^\alpha| = |x_{t+1}^\alpha| < x_{th1} \quad or \quad |k_t^\alpha| = |x_{t+1}^\alpha| > x_{th2} \tag{S62b}$$

$$|z'^\alpha_{t+1}| > r_1 \quad and \quad t > 0 \tag{S62c}$$

If any of the conditions given in Eq. S62a, S62b, or S62c is true, the update process is terminated and a restart is initiated. In this study, the thresholds were set as follows: $b_{th} = 10^{-2}$, $\varepsilon_1 = 10^{-7}$ hartree, $x_{th1} = 0.1$, $x_{th2} = 1.2$, $r_1 = 1$. A restart is also triggered when the upper limit of the number of iterations, max_cycle, is reached. In this study, max_cycle=20.

## Section S2. Details of SBCI2

### S2.1 Derivation of the updating rule for SBCI2

Let us consider the task of determining the lowest $n$ eigenvalues and corresponding eigenvectors of the electronic Hamiltonian $H$ of a molecule. To seek the $\alpha$-th eigenvector and eigenvalue, both state $\alpha$ and state $\alpha + 1$ are concurrently updated within the SBCI2 framework. We introduce the classical Hamiltonian corresponding to these two states as

$$H_{\text{SBCI2}} = \sum_{i=\alpha}^{\alpha+1}\sum_{j=\alpha}^{\alpha+1} \frac{b_t^{i,j}}{2} {y'}_t^{iT} M^{-1} {y'}_t^j + \sum_{i=\alpha}^{\alpha+1} \frac{c_t^{i,i}}{2} \frac{x_t^{iT} H x_t^i}{x_t^{iT} x_t^i}$$
$$+ \sum_{i=\alpha}^{\alpha+1}\sum_{j=\alpha(i\neq j)}^{\alpha+1} \left( \frac{c_t^{i,j}}{2} x_t^{iT} H x_t^j + \frac{d_t^{i,j}}{2} x_t^{iT} x_t^j + \frac{a_t^{i,j}}{2} x_t^{iT} M x_t^j \right) \quad (12)$$

We regard $x_t^\alpha$ and $x_t^{\alpha+1}$ as trial vectors for the $\alpha$-th and $(\alpha+1)$-th eigenvectors, respectively, and ${y'}_t^\alpha$ and ${y'}_t^{\alpha+1}$ are the momentum vectors of this classical system at time $t$. $a_t^{i,j}, b_t^{i,j}, c_t^{i,j}$, and $d_t^{i,j}$ are time-dependent parameters. In Eq. 12, the $\frac{c_t^{i,j}}{2} x_t^{iT} H x_t^j$ term represents the orthogonality condition between $x_t^i$ and $x_t^j$ as eigenvectors of $H$. The $\frac{d_t^{i,j}}{2} x_t^{iT} x_t^j$ term represents the orthogonality condition based on the inner product between $x_t^i$ and $x_t^j$. The $\frac{a_t^{i,j}}{2} x_t^{iT} M x_t^j$ term represents the orthogonality condition based on the inner product between $x_t^i$ and $x_t^j$ scaled by the mass matrix $M$. The Hamilton's equations of motion for this classical system are given by

$$\dot{x}_t^\alpha = \nabla_{{y'}_t^\alpha} H_{\text{SBCI2}} = b_t^{\alpha,\alpha} M^{-1} {y'}_t^\alpha + b_t^{\alpha,\alpha+1} M^{-1} {y'}_t^{\alpha+1} \quad (S63a)$$
$$\dot{x}_t^{\alpha+1} = \nabla_{{y'}_t^{\alpha+1}} H_{\text{SBCI2}} = b_t^{\alpha+1,\alpha+1} M^{-1} {y'}_t^{\alpha+1} + b_t^{\alpha+1,\alpha} M^{-1} {y'}_t^\alpha \quad (S63b)$$
$$\dot{y'}_t^\alpha = -\nabla_{x_t^\alpha} H_{\text{SBCI2}} = -c_t^{\alpha,\alpha} {z'}_t^\alpha - c_t^{\alpha,\alpha+1} H x_t^{\alpha+1} - d_t^{\alpha,\alpha+1} x_t^{\alpha+1} - a_t^{\alpha,\alpha+1} M x_t^{\alpha+1} \quad (S64a)$$
$$\dot{y'}_t^{\alpha+1} = -\nabla_{x_t^{\alpha+1}} H_{\text{SBCI2}} = -c_t^{\alpha+1,\alpha+1} {z'}_t^{\alpha+1} - c_t^{\alpha+1,\alpha} H x_t^\alpha - d_t^{\alpha+1,\alpha} x_t^\alpha - a_t^{\alpha+1,\alpha} M x_t^\alpha \quad (S64b)$$

where dots denote time derivatives, and ${z'}_t^\alpha$ is the residual vector defined as

$${z'}_t^\alpha = \frac{(H - E_t^\alpha I) x_t^\alpha}{x_t^{\alpha T} x_t^\alpha} \quad (4)$$

$E_t^\alpha$ is the Rayleigh quotient defined as

$$E_t^\alpha = \frac{x_t^{\alpha T} H x_t^\alpha}{x_t^{\alpha T} x_t^\alpha} \quad (5)$$

and $I$ is the unit matrix. Introducing $y_t^\alpha$ and $z_t^\alpha$ as

$$y_t^\alpha = M^{-1} {y'}_t^\alpha \quad (6)$$
$$z_t^\alpha = M^{-1} {z'}_t^\alpha \quad (7)$$

Eqs. S63a and S63b are rewritten as

$$\dot{x}_t^\alpha = b_t^{\alpha,\alpha} y_t^\alpha + b_t^{\alpha,\alpha+1} y_t^{\alpha+1} \quad (S65a)$$
$$\dot{x}_t^{\alpha+1} = b_t^{\alpha+1,\alpha+1} y_t^{\alpha+1} + b_t^{\alpha+1,\alpha} y_t^\alpha \quad (S65b)$$

Eq. S64a is rewritten as

$$\dot{y}_t^\alpha = -c_t^{\alpha,\alpha} z_t^\alpha - c_t^{\alpha,\alpha+1} M^{-1} H x_t^{\alpha+1} - d_t^{\alpha,\alpha+1} M^{-1} x_t^{\alpha+1} - a_t^{\alpha,\alpha+1} x_t^{\alpha+1} \quad (S66)$$

Here, we can write
$$M^{-1}Hx_t^{\alpha+1} = (x_t^{\alpha+1T}x_t^{\alpha+1})z_t^{\alpha+1} + E_t^{\alpha+1}M^{-1}x_t^{\alpha+1} \tag{S67}$$
Substituting Eq. S67 into Eq. S66 yields
$$\dot{y}_t^\alpha = -c_t^{\alpha,\alpha}z_t^\alpha - c_t^{\alpha,\alpha+1}(x_t^{\alpha+1T}x_t^\alpha)z_t^{\alpha+1} - (c_t^{\alpha,\alpha+1}E_t^{\alpha+1} + d_t^{\alpha,\alpha+1})M^{-1}x_t^{\alpha+1}$$
$$-a_t^{\alpha,\alpha+1}x_t^{\alpha+1} \tag{S68}$$
Now, we set
$$d_t^{\alpha,\alpha+1} = -c_t^{\alpha,\alpha+1}E_t^{\alpha+1} \tag{S69}$$
Substituting Eq. S69 into Eq. S68 yields
$$\dot{y}_t^\alpha = -c_t^{\alpha,\alpha}z_t^\alpha - c_t^{\alpha,\alpha+1}(x_t^{\alpha+1T}x_t^{\alpha+1})z_t^{\alpha+1} - a_t^{\alpha,\alpha+1}x_t^{\alpha+1} \tag{S70}$$
Here, by redefining $c_t^{\alpha,\alpha+1}(x_t^{\alpha+1T}x_t^\alpha)$ as $c_t^{\alpha,\alpha+1}$, and $a_t^{\alpha,\alpha+1}$ as $-a_t^{\alpha,\alpha+1}$, Eq. S70 is rewritten as
$$\dot{y}_t^\alpha = -c_t^{\alpha,\alpha}z_t^\alpha - c_t^{\alpha,\alpha+1}z_t^{\alpha+1} + a_t^{\alpha,\alpha+1}x_t^{\alpha+1} \tag{S71a}$$
Similarly, Eq. S64b can be rewritten as
$$\dot{y}_t^{\alpha+1} = -c_t^{\alpha+1,\alpha+1}z_t^{\alpha+1} - c_t^{\alpha+1,\alpha}z_t^\alpha + a_t^{\alpha+1,\alpha}x_t^\alpha \tag{S71b}$$
We solve Eqs. S65a S65b S71a and S71b by the symplectic Euler method, where time is discretized with a time step $\Delta_t = 1$. The updating rule is as follows
$$y_{t+1}^\alpha = y_t^\alpha - c_t^{\alpha,\alpha}z_t^\alpha - c_t^{\alpha,\alpha+1}z_t^{\alpha+1} + a_t^{\alpha,\alpha+1}x_t^{\alpha+1} \tag{S72a}$$
$$y_{t+1}^{\alpha+1} = y_t^{\alpha+1} - c_t^{\alpha+1,\alpha+1}z_t^{\alpha+1} - c_t^{\alpha+1,\alpha}z_t^\alpha + a_t^{\alpha+1,\alpha}x_t^\alpha \tag{S72b}$$
$$x_{t+1}^\alpha = x_t^\alpha + b_t^{\alpha,\alpha}y_{t+1}^\alpha + b_t^{\alpha,\alpha+1}y_{t+1}^{\alpha+1} \tag{S73a}$$
$$x_{t+1}^{\alpha+1} = x_t^{\alpha+1} + b_t^{\alpha+1,\alpha+1}y_{t+1}^{\alpha+1} + b_t^{\alpha+1,\alpha}y_{t+1}^\alpha \tag{S73b}$$
updating rule Eqs. S72a S72b S73a S73b are rewritten as
$$y_{t+1} = y_t - C_t z_t + A_t x_t \tag{13}$$
$$x_{t+1} = x_t + B_t y_{t+1} \tag{14}$$
where
$$x_t = \begin{pmatrix} x_t^\alpha \\ x_t^{\alpha+1} \end{pmatrix}, \quad y_t = \begin{pmatrix} y_t^\alpha \\ y_t^{\alpha+1} \end{pmatrix}, \quad z_t = \begin{pmatrix} z_t^\alpha \\ z_t^{\alpha+1} \end{pmatrix} \tag{15}$$
and the real matrices
$$A_t = \begin{pmatrix} 0 & a_t^{\alpha,\alpha+1} \\ a_t^{\alpha+1,\alpha} & 0 \end{pmatrix}, B_t = \begin{pmatrix} b_t^{\alpha,\alpha} & b_t^{\alpha,\alpha+1} \\ b_t^{\alpha+1,\alpha} & b_t^{\alpha+1,\alpha+1} \end{pmatrix}, C_t = \begin{pmatrix} c_t^{\alpha,\alpha} & c_t^{\alpha,\alpha+1} \\ c_t^{\alpha+1,\alpha} & c_t^{\alpha+1,\alpha+1} \end{pmatrix} \tag{16}$$
are defined.

We variationally determine the matrices $A_t$, $B_t$, and $C_t$ such that $x_{t+1}^\alpha$ and $x_{t+1}^{\alpha+1}$ are parallel to the lowest-eigenvalue and the next-lowest-eigenvalue eigenvectors, respectively, of $H$ in the subspace $\{x_t^\alpha, x_t^{\alpha+1}, y_t^\alpha, y_t^{\alpha+1}, z_t^\alpha, z_t^{\alpha+1}\}$, which can be done by diagonalizing the 6 × 6 matrix obtained by representing $H$ using the six orthonormal basis vectors derived from the canonical orthonormalization (*52*) of $x_t^\alpha, x_t^{\alpha+1}, y_t^\alpha, y_t^{\alpha+1}, z_t^\alpha$, and $z_t^{\alpha+1}$. The obtained lowest and the second lowest eigenvalues are $E_{t+1}^\alpha$ and $E_{t+1}^{\alpha+1}$, respectively. The details are described in sections S2.2 and S2.3.

## S2.2 Initial trial vectors of SBCI2

To construct the initial trial vectors, we select $n$ Slater determinants corresponding to the smallest diagonal elements of $H$, and prepare the corresponding vectors $x_{-1}^i$ ($i = 0, \ldots, n-1$). These vectors are then subjected to Gram-Schmidt orthonormalization to verify their linear independence. If the norm of a vector, after subtracting its projection onto the already selected basis, falls below $1.0 \times 10^{-14}$, the corresponding Slater determinant is excluded. We prepare

$$\boldsymbol{X}^i_{-1} = H\boldsymbol{x}^i_{-1} \tag{S74}$$

Using the eigenvectors $U_0$ obtained from the diagonalization of $H$ within the subspace $\{\boldsymbol{x}^i_{-1}\}$, we prepare

$$\boldsymbol{x}^\alpha_0 = \sum_i U^{i,\alpha}_0 \boldsymbol{x}^i_{-1} \tag{S75}$$

for initial trial vectors, and we prepare

$$\boldsymbol{X}^\alpha_0 = H\boldsymbol{x}^\alpha_0 = \sum_i U^{i,\alpha}_0 \boldsymbol{X}^i_{-1} \tag{S76}$$

We also store the corresponding eigenvalues $\{E^i_0\}$.

When we compute the ground state, we employ $\boldsymbol{x}^0_0$ and $\boldsymbol{x}^1_0$ obtained by Eq. S75 as the initial trial vectors. When we compute the excited state $\alpha$, we employ normalized $\boldsymbol{x}^\alpha_{t+1}$, obtained when the state $\alpha - 1$ has converged, as the initial trial vector for state $\alpha$. We redefine it as $\boldsymbol{x}^\alpha_0$. This vector is already orthogonal to the previously determined vectors $\boldsymbol{x}^i_c (i = 0 \sim \alpha - 1)$. For the excited state $\alpha + 1$, we orthogonalize $\boldsymbol{x}^{\alpha+1}_0$ obtained by Eq. S75 against the previously determined vectors $\boldsymbol{x}^i_c (i = 0, ..., \alpha - 1)$ and $\boldsymbol{x}^\alpha_0$, then use it as the initial trial vector. We redefine it as $\boldsymbol{x}^{\alpha+1}_0$.

## S2.3 The first update step of SBCI2

Assume that initial trial vectors $\boldsymbol{x}^\alpha_0, \boldsymbol{x}^{\alpha+1}_0, \boldsymbol{X}^\alpha_0$, and $\boldsymbol{X}^{\alpha+1}_0$ have been obtained. Initially, we set $\boldsymbol{y}_0 = \boldsymbol{0}$ and $B_0 = I$. To orthogonalize $\boldsymbol{x}^\alpha_0$ and $\boldsymbol{x}^{\alpha+1}_0$ against the previously obtained lower-state vectors $\boldsymbol{x}^i_c (i = 0, ..., \alpha - 1)$, we prepare

$$\boldsymbol{x}^{\alpha\prime}_0 = \boldsymbol{x}^\alpha_0 - \sum_{i=0}^{\alpha-1} \boldsymbol{x}^i_c (\boldsymbol{x}^{iT}_c \cdot \boldsymbol{x}^\alpha_0) \tag{S77a}$$

$$\boldsymbol{x}^{\alpha+1\prime}_0 = \boldsymbol{x}^{\alpha+1}_0 - \sum_{i=0}^{\alpha-1} \boldsymbol{x}^i_c (\boldsymbol{x}^{iT}_c \cdot \boldsymbol{x}^{\alpha+1}_0) \tag{S77b}$$

$$\boldsymbol{X}^{\alpha\prime}_0 = \boldsymbol{X}^\alpha_0 - \sum_{i=0}^{\alpha-1} E^i_c \boldsymbol{x}^i_c (\boldsymbol{x}^{iT}_c \cdot \boldsymbol{x}^\alpha_0) \tag{S78a}$$

$$\boldsymbol{X}^{\alpha+1\prime}_0 = \boldsymbol{X}^{\alpha+1}_0 - \sum_{i=0}^{\alpha-1} E^i_c \boldsymbol{x}^i_c (\boldsymbol{x}^{iT}_c \cdot \boldsymbol{x}^{\alpha+1}_0) \tag{S78b}$$

where $E^i_c$ is the eigenvalue for the state $i$. Normalizing $\boldsymbol{x}^{\alpha\prime}_0$ and $\boldsymbol{x}^{\alpha+1\prime}_0$, we prepare

$$\boldsymbol{x}^\alpha_0 = \frac{\boldsymbol{x}^{\alpha\prime}_0}{\sqrt{\boldsymbol{x}^{\alpha\prime T}_0 \cdot \boldsymbol{x}^{\alpha\prime}_0}} \tag{S79a}$$

$$\boldsymbol{x}^{\alpha+1}_0 = \frac{\boldsymbol{x}^{\alpha+1\prime}_0}{\sqrt{\boldsymbol{x}^{\alpha+1\prime T}_0 \cdot \boldsymbol{x}^{\alpha+1\prime}_0}} \tag{S79b}$$

$$X_0^\alpha = \frac{X_0^{\alpha'}}{\sqrt{x_0^{\alpha'T} \cdot x_0^{\alpha'}}} \tag{S80a}$$

$$X_0^{\alpha+1} = \frac{X_0^{\alpha+1'}}{\sqrt{x_0^{\alpha*1'T} \cdot x_0^{\alpha*1'}}} \tag{S80b}$$

A preconditioned residual vector is constructed, and it is orthogonalized against the previously obtained lower-state vectors as follows:

$$z_0^\alpha = \left(1 - \sum_{i=0}^{\alpha-1} x_c^i x_c^{iT}\right)(D - E^0 I)^{-1}(H - E_0^\alpha I)x_0^\alpha \tag{S81a}$$

$$z_0^{\alpha+1} = \left(1 - \sum_{i=0}^{\alpha-1} x_c^i x_c^{iT}\right)(D - E^0 I)^{-1}(H - E_0^{\alpha+1} I)x_0^{\alpha+1} \tag{S81b}$$

We prepare

$$Z_0^\alpha = H z_0^\alpha \tag{S82a}$$
$$Z_0^{\alpha+1} = H z_0^{\alpha+1} \tag{S82b}$$

Next, the vectors $x_0^\alpha, x_0^{\alpha+1}, z_0^\alpha$, and $z_0^{\alpha+1}$ are transformed into the orthonormal basis $e_0$ via canonical orthogonalization (*52*). This is carried out as follows. First, all pairwise inner products between $x_0^\alpha, x_0^{\alpha+1}, z_0^\alpha$, and $z_0^{\alpha+1}$ are computed to obtain the overlap matrix $S$. The matrix $S$ is then fully diagonalized to yield the diagonal matrix $s$, containing the eigenvalues, and the matrix $U$, whose columns are the corresponding eigenvectors. At this stage, eigenvalues that are negative or less than $10^{-14}$ are excluded. From this, the transformation matrix to the orthonormal basis is computed as $P = Us^{-1/2}$. Consequently, the orthonormal basis $e_0$ can be expressed as follows:

$$e_0 = (x_0^\alpha \quad x_0^{\alpha+1} \quad z_0^\alpha \quad z_0^{\alpha+1})P \tag{S83}$$

The matrix representing the Hamiltonian $H$ in the subspace spanned by $e_0$ can be written as

$$e_0^T H e_0 = P^T(x_0^\alpha \quad x_0^{\alpha+1} \quad z_0^\alpha \quad z_0^{\alpha+1})^T H (x_0^\alpha \quad x_0^{\alpha+1} \quad z_0^\alpha \quad z_0^{\alpha+1})P \tag{S84}$$

Diagonalizing this matrix, the eigenvectors can be expressed as follows:

$$v = e_0 V = (x_0^\alpha \quad x_0^{\alpha+1} \quad z_0^\alpha \quad z_0^{\alpha+1})PV \tag{S85}$$

Defining the matrix

$$V' = PV = \begin{pmatrix} V'_{0,0} & V'_{0,1} & V'_{0,2} & V'_{0,3} \\ V'_{1,0} & V'_{1,1} & V'_{1,2} & V'_{1,3} \\ V'_{2,0} & V'_{2,1} & V'_{2,2} & V'_{2,3} \\ V'_{3,0} & V'_{3,1} & V'_{3,2} & V'_{3,3} \end{pmatrix} \tag{S86}$$

we can write Eq. S85 as

$$v = (v^0 \quad v^1 \quad v^2 \quad v^3) = (x_0^\alpha \quad x_0^{\alpha+1} \quad z_0^\alpha \quad z_0^{\alpha+1})V' \tag{S87}$$

On the other hand, according to Eqs. 13 and 14, the update formula is given as follows:

$$x_1 = (I + A_0)x_0 - C_0 z_0 \tag{S88}$$

which can be expressed in terms of elements as follows:

$$\begin{pmatrix} x_1^\alpha \\ x_1^{\alpha+1} \end{pmatrix} = \begin{pmatrix} x_0^\alpha + a_0^{\alpha,\alpha+1} x_0^{\alpha+1} - c_0^{\alpha,\alpha} z_0^\alpha - c_0^{\alpha,\alpha+1} z_0^{\alpha+1} \\ a_0^{\alpha+1,\alpha} x_0^\alpha + x_0^{\alpha+1} - c_0^{\alpha+1,\alpha} z_0^\alpha - c_0^{\alpha+1,\alpha+1} z_0^{\alpha+1} \end{pmatrix} \quad \text{(S89)}$$

The condition that $x_1^\alpha$ and $x_1^{\alpha+1}$ should be respectively parallel to $v^0$ and $v^1$ can be expressed as follows:

$$\begin{pmatrix} x_1^\alpha \\ x_1^{\alpha+1} \end{pmatrix} = \begin{pmatrix} k_0 v^0 \\ k_1 v^1 \end{pmatrix} \quad \text{(S90)}$$

By substituting Eq. S89 into the left-hand side of Eq. S90, and using Eq. S87 in the right-hand side of Eq. S90, the following identity is obtained:

$$\begin{pmatrix} x_0^\alpha + a_0^{\alpha,\alpha+1} x_0^{\alpha+1} - c_0^{\alpha,\alpha} z_0^\alpha - c_0^{\alpha,\alpha+1} z_0^{\alpha+1} \\ a_0^{\alpha+1,\alpha} x_0^\alpha + x_0^{\alpha+1} - c_0^{\alpha+1,\alpha} z_0^\alpha - c_0^{\alpha+1,\alpha+1} z_0^{\alpha+1} \end{pmatrix} = \begin{pmatrix} k_0 \left( x_0^\alpha V'_{0,0} + x_0^{\alpha+1} V'_{1,0} + z_0^\alpha V'_{2,0} + z_0^{\alpha+1} V'_{3,0} \right) \\ k_1 \left( x_0^\alpha V'_{0,1} + x_0^{\alpha+1} V'_{1,1} + z_0^\alpha V'_{2,1} + z_0^{\alpha+1} V'_{3,1} \right) \end{pmatrix} \quad \text{(S91)}$$

From the equality of both sides of Eq. S91, the following equations are obtained:

$$k_0 V'_{0,0} = 1, \quad k_0 V'_{1,0} = a_0^{\alpha,\alpha+1}, \quad k_0 V'_{2,0} = -c_0^{\alpha,\alpha}, \quad k_0 V'_{3,0} = -c_0^{\alpha,\alpha+1} \quad \text{(S92a)}$$

$$k_1 V'_{0,1} = a_0^{\alpha+1,\alpha}, \quad k_1 V'_{1,1} = 1, \quad k_1 V'_{2,1} = -c_0^{\alpha+1,\alpha}, \quad k_1 V'_{3,1} = -c_0^{\alpha+1,\alpha+1} \quad \text{(S92b)}$$

Using Eqs. S92a and S92b, we obtain

$$k_0 = \frac{1}{V'_{0,0}}, \quad a_0^{\alpha,\alpha+1} = \frac{V'_{1,0}}{V'_{0,0}}, \quad c_0^{\alpha,\alpha} = -\frac{V'_{2,0}}{V'_{0,0}}, \quad c_0^{\alpha,\alpha+1} = -\frac{V'_{3,0}}{V'_{0,0}} \quad \text{(S93a)}$$

$$k_1 = \frac{1}{V'_{1,1}}, \quad a_0^{\alpha+1,\alpha} = \frac{V'_{0,1}}{V'_{1,1}}, \quad c_0^{\alpha+1,\alpha+1} = -\frac{V'_{3,1}}{V'_{1,1}}, \quad c_0^{\alpha+1,\alpha} = -\frac{V'_{2,1}}{V'_{1,1}} \quad \text{(S93b)}$$

Now, we can perform updating as follows:

$$y_1^\alpha = -c_0^{\alpha,\alpha} z_0^\alpha - c_0^{\alpha,\alpha+1} z_0^{\alpha+1} + a_0^{\alpha,\alpha+1} x_0^{\alpha+1} \quad \text{(S94a)}$$

$$y_1^{\alpha+1} = -c_0^{\alpha+1,\alpha} z_0^\alpha - c_0^{\alpha+1,\alpha+1} z_0^{\alpha+1} + a_0^{\alpha+1,\alpha} x_0^\alpha \quad \text{(S94b)}$$

$$Y_1^\alpha = H y_1^\alpha = -c_0^{\alpha,\alpha} Z_0^\alpha - c_0^{\alpha,\alpha+1} Z_0^{\alpha+1} + a_0^{\alpha,\alpha+1} X_0^{\alpha+1} \quad \text{(S95a)}$$

$$Y_1^{\alpha+1} = H y_1^{\alpha+1} = -c_0^{\alpha+1,\alpha} Z_0^\alpha - c_0^{\alpha+1,\alpha+1} Z_0^{\alpha+1} + a_0^{\alpha+1,\alpha} X_0^\alpha \quad \text{(S95b)}$$

$$x_1^\alpha = x_0^\alpha + y_1^\alpha \quad \text{(S96a)}$$

$$x_1^{\alpha+1} = x_0^{\alpha+1} + y_1^{\alpha+1} \quad \text{(S96b)}$$

$$X_1^\alpha = H x_1^\alpha = X_0^\alpha + Y_1^\alpha \quad \text{(S97a)}$$

$$X_1^{\alpha+1} = H x_1^{\alpha+1} = X_0^{\alpha+1} + Y_1^{\alpha+1} \quad \text{(S97b)}$$

The residual vector is then constructed as follows:

$$z'^\alpha_1 = \frac{1}{k_0}(H - E_1^\alpha I) x_1^\alpha \quad \text{(S98)}$$

and if the convergence criterion

$$|E_1^\alpha - E_0^\alpha| < \varepsilon_0 \quad \text{and} \quad |z'^\alpha_1| < r_0 \quad \text{(S99)}$$

is satisfied, then $x_1^\alpha$ is normalized and stored as $x_c^\alpha$. $E_1^\alpha$ is stored as $E_c^\alpha$, and the solution process for state $\alpha$ is terminated. If convergence is not achieved, the residual vector for state $\alpha + 1$ is constructed as

$$z'^{\alpha+1}_1 = \frac{1}{k_1}(H - E^{\alpha+1}_1 I)x^{\alpha+1}_1 \tag{S100}$$

and the procedure continues. In this study, if frozen core=No, $\varepsilon_0 = 10^{-10}$ hartree and $r_0 = 10^{-5}$. If frozen core=Yes, $\varepsilon_0 = 10^{-8}$ hartree and $r_0 = 10^{-4}$.

### S2.4 The following update step of SBCI2

Consider time step $t(\geq 1)$. Assume that $x^\alpha_t, X^\alpha_t(= Hx^\alpha_t), y^\alpha_t, Y^\alpha_t(= Hy^\alpha_t), x^{\alpha+1}_t, X^{\alpha+1}_t(= Hx^{\alpha+1}_t), y^{\alpha+1}_t, Y^{\alpha+1}_t(= Hy^{\alpha+1}_t), z'^\alpha_t$, and $z'^{\alpha+1}_t$ have been obtained. A preconditioned residual vector is constructed, and it is orthogonalized against the previously obtained lower-state vectors as follows:

$$z^\alpha_t = \left(1 - \sum_{i=0}^{\alpha-1} x^i_c x^{iT}_c\right)(D - E^0 I)^{-1} z'^\alpha_t \tag{S101a}$$

$$z^{\alpha+1}_t = \left(1 - \sum_{i=0}^{\alpha-1} x^i_c x^{iT}_c\right)(D - E^0 I)^{-1} z'^{\alpha+1}_t \tag{S101b}$$

We prepare
$$Z^\alpha_t = Hz^\alpha_t \tag{S102a}$$
$$Z^{\alpha+1}_t = Hz^{\alpha+1}_t \tag{S102b}$$

Next, the vectors $x^\alpha_t, x^{\alpha+1}_t, y^\alpha_t, y^{\alpha+1}_t, z^\alpha_t$, and $z^{\alpha+1}_t$ are transformed into the orthonormal basis $e_t$ via canonical orthogonalization (52). This is carried out as follows. First, all pairwise inner products between $x^\alpha_t, x^{\alpha+1}_t, y^\alpha_t, y^{\alpha+1}_t, z^\alpha_t$, and $z^{\alpha+1}_t$ are computed to obtain the overlap matrix $S$. The matrix $S$ is then fully diagonalized to yield the diagonal matrix $s$, containing the eigenvalues, and the matrix $U$, whose columns are the corresponding eigenvectors. At this stage, eigenvalues that are negative or less than $10^{-14}$ are excluded. From this, the transformation matrix to the orthonormal basis is computed as $P = Us^{-1/2}$. Consequently, the orthonormal basis $e_t$ can be expressed as follows:

$$e_t = (x^\alpha_t \quad x^{\alpha+1}_t \quad y^\alpha_t \quad y^{\alpha+1}_t \quad z^\alpha_t \quad z^{\alpha+1}_t)\, P \;. \tag{S103}$$

The matrix representing the Hamiltonian $H$ in the subspace spanned by $e_t$ can be written as

$$e_t^T H e_t$$
$$= P^T (x^\alpha_t \quad x^{\alpha+1}_t \quad y^\alpha_t \quad y^{\alpha+1}_t \quad z^\alpha_t \quad z^{\alpha+1}_t)^T H (x^\alpha_t \quad x^{\alpha+1}_t \quad y^\alpha_t \quad y^{\alpha+1}_t \quad z^\alpha_t \quad z^{\alpha+1}_t) P \tag{S104}$$

Diagonalizing this matrix, the eigenvectors can be expressed as follows:

$$v = e_t V = (x^\alpha_t \quad x^{\alpha+1}_t \quad y^\alpha_t \quad y^{\alpha+1}_t \quad z^\alpha_t \quad z^{\alpha+1}_t) PV \tag{S105}$$

Defining the matrix

$$V' = PV = \begin{pmatrix} V'_{0,0} & V'_{0,1} & V'_{0,2} & V'_{0,3} & V'_{0,4} & V'_{0,5} \\ V'_{1,0} & V'_{1,1} & V'_{1,2} & V'_{1,3} & V'_{1,4} & V'_{1,5} \\ V'_{2,0} & V'_{2,1} & V'_{2,2} & V'_{2,3} & V'_{2,4} & V'_{2,5} \\ V'_{3,0} & V'_{3,1} & V'_{3,2} & V'_{3,3} & V'_{3,4} & V'_{3,5} \\ V'_{4,0} & V'_{4,1} & V'_{4,2} & V'_{4,3} & V'_{4,4} & V'_{4,5} \\ V'_{5,0} & V'_{5,1} & V'_{5,2} & V'_{5,3} & V'_{5,4} & V'_{5,5} \end{pmatrix} \tag{S106}$$

we can write Eq. S105 as

$$\boldsymbol{v} = (\boldsymbol{v}^0 \quad \boldsymbol{v}^1 \quad \boldsymbol{v}^2 \quad \boldsymbol{v}^3 \quad \boldsymbol{v}^4 \quad \boldsymbol{v}^5) = (\boldsymbol{x}_t^\alpha \quad \boldsymbol{x}_t^{\alpha+1} \quad \boldsymbol{y}_t^\alpha \quad \boldsymbol{y}_t^{\alpha+1} \quad \boldsymbol{z}_t^\alpha \quad \boldsymbol{z}_t^{\alpha+1}) V' \quad (S107)$$

On the other hand, according to Eqs. 13 and 14, the update formula is given as follows:

$$\boldsymbol{x}_{t+1} = (I + B_t A_t)\boldsymbol{x}_t + B_t \boldsymbol{y}_t - B_t C_t \boldsymbol{z}_t \quad (S108)$$

which can be expressed in terms of elements as follows:

$$\begin{pmatrix} \boldsymbol{x}_{t+1}^\alpha \\ \boldsymbol{x}_{t+1}^{\alpha+1} \end{pmatrix} = \begin{pmatrix} (1 + b_t^{\alpha,\alpha+1} a_t^{\alpha+1,\alpha})\boldsymbol{x}_t^\alpha + b_t^{\alpha,\alpha} a_t^{\alpha,\alpha+1} \boldsymbol{x}_t^{\alpha+1} \\ b_t^{\alpha+1,\alpha+1} a_t^{\alpha+1,\alpha} \boldsymbol{x}_t^\alpha + (1 + b_t^{\alpha+1,\alpha} a_t^{\alpha,\alpha+1})\boldsymbol{x}_t^{\alpha+1} \end{pmatrix}$$
$$+ \begin{pmatrix} b_t^{\alpha,\alpha} \boldsymbol{y}_t^\alpha + b_t^{\alpha,\alpha+1} \boldsymbol{y}_t^{\alpha+1} \\ b_t^{\alpha+1,\alpha} \boldsymbol{y}_t^\alpha + b_t^{\alpha+1,\alpha+1} \boldsymbol{y}_t^{\alpha+1} \end{pmatrix} - B_t C_t \begin{pmatrix} \boldsymbol{z}_t^\alpha \\ \boldsymbol{z}_t^{\alpha+1} \end{pmatrix} \quad (S109)$$

The condition that $\boldsymbol{x}_{t+1}^\alpha$ and $\boldsymbol{x}_{t+1}^{\alpha+1}$ should be respectively parallel to $\boldsymbol{v}^0$ and $\boldsymbol{v}^1$ can be expressed as follows:

$$\begin{pmatrix} \boldsymbol{x}_{t+1}^\alpha \\ \boldsymbol{x}_{t+1}^{\alpha+1} \end{pmatrix} = \begin{pmatrix} k_0 \boldsymbol{v}^0 \\ k_1 \boldsymbol{v}^1 \end{pmatrix} \quad (S110)$$

By substituting Eq. S109 into the left-hand side of Eq. S110, and using Eq. S107 in the right-hand side of Eq. S110, the following identity is obtained:

$$\begin{pmatrix} (1 + b_t^{\alpha,\alpha+1} a_t^{\alpha+1,\alpha})\boldsymbol{x}_t^\alpha + b_t^{\alpha,\alpha} a_t^{\alpha,\alpha+1} \boldsymbol{x}_t^{\alpha+1} \\ b_t^{\alpha+1,\alpha+1} a_t^{\alpha+1,\alpha} \boldsymbol{x}_t^\alpha + (1 + b_t^{\alpha+1,\alpha} a_t^{\alpha,\alpha+1})\boldsymbol{x}_t^{\alpha+1} \end{pmatrix}$$
$$+ \begin{pmatrix} b_t^{\alpha,\alpha} \boldsymbol{y}_t^\alpha + b_t^{\alpha,\alpha+1} \boldsymbol{y}_t^{\alpha+1} \\ b_t^{\alpha+1,\alpha} \boldsymbol{y}_t^\alpha + b_t^{\alpha+1,\alpha+1} \boldsymbol{y}_t^{\alpha+1} \end{pmatrix} - B_t C_t \begin{pmatrix} \boldsymbol{z}_t^\alpha \\ \boldsymbol{z}_t^{\alpha+1} \end{pmatrix}$$
$$= \begin{pmatrix} k_0(\boldsymbol{x}_t^\alpha V'_{0,0} + \boldsymbol{x}_t^{\alpha+1} V'_{1,0} + \boldsymbol{y}_t^\alpha V'_{2,0} + \boldsymbol{y}_t^{\alpha+1} V'_{3,0} + \boldsymbol{z}_t^\alpha V'_{4,0} + \boldsymbol{z}_t^{\alpha+1} V'_{5,0}) \\ k_1(\boldsymbol{x}_t^\alpha V'_{0,1} + \boldsymbol{x}_t^{\alpha+1} V'_{1,1} + \boldsymbol{y}_t^\alpha V'_{2,1} + \boldsymbol{y}_t^{\alpha+1} V'_{3,1} + \boldsymbol{z}_t^\alpha V'_{4,1} + \boldsymbol{z}_t^{\alpha+1} V'_{5,1}) \end{pmatrix} \quad (S111)$$

Comparing the coefficients of $\boldsymbol{y}_t^\alpha$ and $\boldsymbol{y}_t^{\alpha+1}$ on both sides of Eq. S111, the following equations obtained:

$$b_t^{\alpha,\alpha} = k_0 V'_{2,0}, \quad b_t^{\alpha,\alpha+1} = k_0 V'_{3,0}, \quad b_t^{\alpha+1,\alpha} = k_1 V'_{2,1}, \quad b_t^{\alpha+1,\alpha+1} = k_1 V'_{3,1} \quad (S112)$$

Comparing the coefficients of $\boldsymbol{x}_t^\alpha$ and $\boldsymbol{x}_t^{\alpha+1}$ on both sides of Eq. S111, the following equations obtained:

$$1 + b_t^{\alpha,\alpha+1} a_t^{\alpha+1,\alpha} = k_0 V'_{0,0}, \quad b_t^{\alpha,\alpha} a_t^{\alpha,\alpha+1} = k_0 V'_{1,0}$$
$$b_t^{\alpha+1,\alpha+1} a_t^{\alpha+1,\alpha} = k_1 V'_{0,1}, \quad 1 + b_t^{\alpha+1,\alpha} a_t^{\alpha,\alpha+1} = k_1 V'_{1,1} \quad (S113)$$

Substituting Eq. S112 into Eq. S113, we obtain the following equations:

$$1 + k_0 V'_{3,0} a_t^{\alpha+1,\alpha} = k_0 V'_{0,0} \quad (S114a)$$

$$k_0 V'_{2,0} a_t^{\alpha,\alpha+1} = k_0 V'_{1,0} \quad (S114b)$$

$$k_1 V'_{3,1} a_t^{\alpha+1,\alpha} = k_1 V'_{0,1} \quad (S115c)$$

$$1 + k_1 V'_{2,1} a_t^{\alpha,\alpha+1} = k_1 V'_{1,1} \quad (S115d)$$

Using Eqs. S114b and S114c, we obtain

$$a_t^{\alpha,\alpha+1} = \frac{V'_{1,0}}{V'_{2,0}}, \quad a_t^{\alpha+1,\alpha} = \frac{V'_{0,1}}{V'_{3,1}} \quad (S116)$$

Using Eqs. S114a and S114d, we obtain

$$k_0 = \frac{1}{V'_{0,0} - V'_{3,0}a_t^{\alpha+1,\alpha}}, \quad k_1 = \frac{1}{V'_{1,1} - V'_{2,1}a_t^{\alpha,\alpha+1}} \tag{S117}$$

Using Eq. S112 with Eq. S117, we can calculate $b_t^{\alpha,\alpha}$, $b_t^{\alpha+1,\alpha}$, $b_t^{\alpha,\alpha+1}$ and $b_t^{\alpha+1,\alpha+1}$. Comparing the coefficients of $z_t^\alpha$ and $z_t^{\alpha+1}$ on both sides of Eq. S111, we obtain

$$-B_t C_t = \begin{pmatrix} k_0 V'_{4,0} & k_0 V'_{5,0} \\ k_1 V'_{4,1} & k_1 V'_{5,1} \end{pmatrix} \tag{S118}$$

Substituting Eq. S112 into Eq. S118, we obtain

$$-\begin{pmatrix} k_0 V'_{2,0} & k_0 V'_{3,0} \\ k_1 V'_{2,1} & k_1 V'_{3,1} \end{pmatrix} C_t = \begin{pmatrix} k_0 V'_{4,0} & k_0 V'_{5,0} \\ k_1 V'_{4,1} & k_1 V'_{5,1} \end{pmatrix} \tag{S119}$$

We can rewrite Eq. S119 as

$$-\begin{pmatrix} V'_{2,0} & V'_{3,0} \\ V'_{2,1} & V'_{3,1} \end{pmatrix} C_t = \begin{pmatrix} V'_{4,0} & V'_{5,0} \\ V'_{4,1} & V'_{5,1} \end{pmatrix} \tag{S120}$$

Solving Eq. S120 for $C_t$, we obtain

$$C_t = -\begin{pmatrix} V'_{2,0} & V'_{3,0} \\ V'_{2,1} & V'_{3,1} \end{pmatrix}^{-1} \begin{pmatrix} V'_{4,0} & V'_{5,0} \\ V'_{4,1} & V'_{5,1} \end{pmatrix} \tag{S121}$$

Now, we can perform updating as follows:

$$y_{t+1}^\alpha = y_t^\alpha - c_t^{\alpha,\alpha} z_t^\alpha - c_t^{\alpha,\alpha+1} z_t^{\alpha+1} + a_t^{\alpha,\alpha+1} x_t^{\alpha+1} \tag{S122a}$$
$$y_{t+1}^{\alpha+1} = y_t^{\alpha+1} - c_t^{\alpha+1,\alpha} z_t^\alpha - c_t^{\alpha+1,\alpha+1} z_t^{\alpha+1} + a_t^{\alpha+1,\alpha} x_t^\alpha \tag{S122b}$$
$$Y_{t+1}^\alpha = H y_{t+1}^\alpha = Y_t^\alpha - c_t^{\alpha,\alpha} Z_t^\alpha - c_t^{\alpha,\alpha+1} Z_t^{\alpha+1} + a_t^{\alpha,\alpha+1} X_t^{\alpha+1} \tag{S123a}$$
$$Y_{t+1}^{\alpha+1} = H y_{t+1}^{\alpha+1} = Y_t^{\alpha+1} - c_t^{\alpha+1,\alpha} Z_t^\alpha - c_t^{\alpha+1,\alpha+1} Z_t^{\alpha+1} + a_t^{\alpha+1,\alpha} X_t^\alpha \tag{S123b}$$
$$x_{t+1}^\alpha = x_t^\alpha + b_t^{\alpha,\alpha} y_{t+1}^\alpha + b_t^{\alpha,\alpha+1} y_{t+1}^{\alpha+1} \tag{S124a}$$
$$x_{t+1}^{\alpha+1} = x_t^{\alpha+1} + b_t^{\alpha+1,\alpha} y_{t+1}^\alpha + b_t^{\alpha+1,\alpha+1} y_{t+1}^{\alpha+1} \tag{S124b}$$
$$X_{t+1}^\alpha = H x_{t+1}^\alpha = X_t^\alpha + b_t^{\alpha,\alpha} Y_{t+1}^\alpha + b_t^{\alpha,\alpha+1} Y_{t+1}^{\alpha+1} \tag{S125a}$$
$$X_{t+1}^{\alpha+1} = H x_{t+1}^{\alpha+1} = X_t^{\alpha+1} + b_t^{\alpha+1,\alpha} Y_{t+1}^\alpha + b_t^{\alpha+1,\alpha+1} Y_{t+1}^{\alpha+1} \tag{S125b}$$

The residual vector is then constructed as follows:

$$z'^\alpha_{t+1} = \frac{1}{k_0}(H - E_{t+1}^\alpha I) x_{t+1}^\alpha \tag{S126}$$

and if the convergence criterion

$$|E_{t+1}^\alpha - E_t^\alpha| < \varepsilon_0 \quad \text{and} \quad |z'^\alpha_{t+1}| < r_0 \tag{S127}$$

is satisfied, then $x_{t+1}^\alpha$ is normalized and stored as $x_c^\alpha$. $E_{t+1}^\alpha$ is stored as $E_c^\alpha$, and the solution process for state $\alpha$ is terminated. If convergence is not achieved, the residual vector for state $\alpha + 1$ is constructed as

$$z'^{\alpha+1}_{t+1} = \frac{1}{k_1}(H - E_{t+1}^{\alpha+1} I) x_{t+1}^{\alpha+1} \tag{S128}$$

and the procedure continues. In this study, if frozen core=No, $\varepsilon_0 = 10^{-10}$ hartree and $r_0 = 10^{-5}$. If frozen core=Yes, $\varepsilon_0 = 10^{-8}$ hartree and $r_0 = 10^{-4}$.

## S2.5 Restart of SBCI2

If convergence is not achieved under the criteria defined by Eq. S99 or Eq. S127 with Eq. S129, each of the following conditions was evaluated:

$$\left(|b_t^{\alpha,\alpha}| < b_{th} \text{ or } |b_t^{\alpha+1,\alpha+1}| < b_{th}\right) \quad and \quad |E_{t+1}^{\alpha} - E_t^{\alpha}| < \varepsilon_1 \quad \text{(S130a)}$$

$$|x_{t+1}^{\alpha}| < x_{th1} \quad or \quad |x_{t+1}^{\alpha+1}| < x_{th1} \text{ or } |x_{t+1}^{\alpha}| > x_{th2} \text{ or } |x_{t+1}^{\alpha+1}| > x_{th2} \quad \text{(S130b)}$$

$$|z'^{\alpha}_{t+1}| > r_1 \quad and \quad t > 0 \quad \text{(S130c)}$$

If any of the conditions given in Eq. S130a, S130b, or S130c is true, the update process is terminated and a restart is initiated. In this study, the thresholds were set as follows: $b_{th} = 10^{-2}$, $\varepsilon_1 = 10^{-7}$ hartree, $x_{th1} = 0.1$, $x_{th2} = 1.2$, $r_1 = 1$. A restart is also triggered when the upper limit of the number of iterations, max_cycle, is reached. In this study, max_cycle=10.

# Section S3. Instances of FCI in this study

## S3.1 Setup of each instance

| molecule | point group | IRep | $S_z$ | $n$ | geometry | basis set | frozen core | $N_{det}$ |
|---|---|---|---|---|---|---|---|---|
| $N_2^*$ | $D_{\infty h}$ | $A_{1g}$ | 0 | 1 | see S3.2.1 | ccpvdz | Yes CAS (10e, 26o) | 4327008400 |
| $CN^*$ | $C_{2v}$ | $A_1$ | 0 | 1 | see S3.2.2 | ccpvdz | Yes CAS(9e,26o) | 983411000 |
| Ne | $D_{2h}$ | $A_g$ | 0 | 9 | - | see S3.3.1 | No CAS(10e,18o) | 73410624 |
| | | $B_{1g}$ | 0 | 6 | | | | |
| | | $B_{1u}$ | 0 | 3 | | | | |
| $H_2O$ | $C_{2v}$ | $A_1$ | 0 | 4 | see S.3.2.3 | see S.3.3.2 | Yes CAS(8e,29o) | 564110001 |
| | | $A_2$ | 0 | 3 | | | | |
| | | $B_1$ | 0 | 3 | | | | |
| | | $B_2$ | 0 | 5 | | | | |
| HF | $C_{2v}$ | $A_2$ | 1 | 3 | see S3.2.4 | augccpvdz | Yes CAS (8e, 31o) | 763749945 |
| | $C_{\infty v}$ | $A_1$ | 0 | 4 | | | | 990046225 |
| | | $E_{1x}$ | 0 | 5 | | | | |
| | | $E_{2uy}$ | 0 | 4 | | | | |
| | | $A_{1g}$ | 0 | 1 | | | | |
| $N_2$ | $D_{\infty h}$ | $E_{1uy}$ | 0 | 3 | see S3.2.5 | ccpvdz | Yes CAS (10e, 26o) | 4327008400 |
| | | $E_{1gx}$ | 0 | 3 | | | | |
| | | $A_{1u}$ | 0 | 1 | | | | |
| BH | $C_{\infty v}$ | $E_{1x}$ | 0 | 6 | see S3.2.6 | see S3.3.3 | No CAS(6e,45o) | 201356100 |
| | | $A_1$ | 0 | 8 | | | | |
| $C_2$ | $D_{\infty h}$ | $E_{2gx}$ | 0 | 5 | see S3.2.7 | see S3.3.4 | Yes CAS (8e, 34o) | 2150733376 |
| | | $A_{1g}$ | 0 | 3 | | | | |
| | | $E_{1uy}$ | 0 | 3 | | | | |
| | | $E_{1gx}$ | 0 | 3 | | | | |
| | | $A_{1u}$ | 0 | 2 | | | | |
| | $C_{2v}$ | $A_2$ | 1 | 3 | | | | 1665083904 |
| $H_2O^+$ | $C_{2v}$ | $A_1$ | 0.5 | 1 | see S3.2.8 | ccpvdz | No CAS(9e,24o) | 451647504 |
| | | $B_1$ | 0.5 | 1 | | | | |
| | | $B_2$ | 0.5 | 1 | | | | |
| $F_2$ | $D_{\infty h}$ | $A_{1g}$ | 0 | 1 | see S3.2.9 | 6-31g | No CAS(18e,18o) | 2363904400 |

*: Instance for potential energy curve  
$n$ : The number of states  
$N_{det}$ : The number of determinants

### S3.2 Geometry for each instance (in Å)

### S3.2.1 Geometry for the potential energy curve of $N_2$

R(N-N) = 0.793765812, 0.952518975, 1.094338467, 1.111272137, 1.120797327, 1.147256188, 1.173715048, 1.2700253, 1.428778462, 1.587531625, 1.746284787, 1.90503795

These values are obtained by transforming values in ref. (*49*) (in Bohr) to those in Å. These values correspond to those in ref. (*46*) (in Å).

### S3.2.2 Geometry for the potential energy curve of CN

R(C-N) = 0.9, 1.0, 1.0918, 1.1318, 1.1518, 1.1569, 1.1718, 1.1769, 1.1869, 1.1918, 1.1919, 1.1969, 1.2019, 1.2069, 1.2118, 1.2169, 1.2369, 1.2518, 1.30, 1.40, 1.50, 1.60, 1.7577, 2.05065

These values are the same as those in ref. (*48*).

### S.3.2.3 Geometry for $H_2O$

```
O  0.000000   0.000000    0.000000
H  0.000000   0.756690220 -0.585891835
H  0.000000  -0.756690220 -0.585891835
```

This geometry corresponds to that in ref. (*44*).

### S.3.2.4 Geometry for HF

```
H    0.000000   0.000000   0.000
F    0.000000   0.000000   0.917
```

This geometry corresponds to that in ref. (*47*).

### S.3.2.5 Geometry for $N_2$

```
N  0.000000       0.000000    0.000000
N  1.0943384670   0.000000    0.000000
```

This geometry corresponds to that in ref. (*44*).

### S.3.2.6 Geometry for BH

```
B    0.000000   0.000000   0.0
H    0.000000   0.000000   1.2324008
```

This geometry corresponds to that in ref. (*43*).

### S.3.2.7 Geometry for C$_2$

```
C  0.000000     0.000000   0.000000
C  1.242508085  0.000000   0.000000
```

This geometry corresponds to that in ref. (*44*).

### S.3.2.8 Geometry for H$_2$O$^+$

```
O   0.000000    0.000000    -0.004763
H   0.000000    0.801843    -0.560345
H   0.000000   -0.801843    -0.560345
```

This geometry corresponds to that in ref. (*45*).

### S.3.2.9 Geometry for F$_2$

```
F   0.000000   0.000000   0.671
F   0.000000   0.000000  -0.671
```

This geometry corresponds to that in ref. (*3*).

## S3.3 Basis sets in PySCF format

### S3.3.1 Basis set for Ne

```
Ne   S
     1.788000E+04       7.380000E-04      -1.720000E-04       0.000000E+00
     2.683000E+03       5.677000E-03      -1.357000E-03       0.000000E+00
     6.115000E+02       2.888300E-02      -6.737000E-03       0.000000E+00
     1.735000E+02       1.085400E-01      -2.766300E-02       0.000000E+00
     5.664000E+01       2.909070E-01      -7.620800E-02       0.000000E+00
     2.042000E+01       4.483240E-01      -1.752270E-01       0.000000E+00
     7.810000E+00       2.580260E-01      -1.070380E-01       0.000000E+00
     1.653000E+00       1.506300E-02       5.670500E-01       0.000000E+00
     4.869000E-01      -2.100000E-03       5.652160E-01       1.000000E+00
Ne   P
     2.839000E+01       4.608700E-02       0.000000E+00
     6.270000E+00       2.401810E-01       0.000000E+00
     1.695000E+00       5.087440E-01       0.000000E+00
     4.317000E-01       4.556600E-01       1.000000E+00
Ne   D
     2.202000E+00       1.0000000
Ne   S
     4.000000E-02       1.0000000
Ne   P
     3.000000E-02       1.0000000
```

This basis set is the same as that in ref. (*43*). One s and one p function with exponents 0.04 and 0.03 are added to ccpvdz.

**S3.3.2 Basis set for H₂O**

```
O  S
    1.172000E+04     7.100000E-04    -1.600000E-04     0.000000E+00
    1.759000E+03     5.470000E-03    -1.263000E-03     0.000000E+00
    4.008000E+02     2.783700E-02    -6.267000E-03     0.000000E+00
    1.137000E+02     1.048000E-01    -2.571600E-02     0.000000E+00
    3.703000E+01     2.830620E-01    -7.092400E-02     0.000000E+00
    1.327000E+01     4.487190E-01    -1.654110E-01     0.000000E+00
    5.025000E+00     2.709520E-01    -1.169550E-01     0.000000E+00
    1.013000E+00     1.545800E-02     5.573680E-01     0.000000E+00
    3.023000E-01    -2.585000E-03     5.727590E-01     1.000000E+00
O  P
    1.770000E+01     4.301800E-02     0.000000E+00
    3.854000E+00     2.289130E-01     0.000000E+00
    1.046000E+00     5.087280E-01     0.000000E+00
    2.753000E-01     4.605310E-01     1.000000E+00
O  D
    1.185000E+00     1.0000000
O  S
    7.896000E-02     1.0000000
O  P
    6.856000E-02     1.0000000
H  S
    1.301000E+01     1.968500E-02     0.000000E+00
    1.962000E+00     1.379770E-01     0.000000E+00
    4.446000E-01     4.781480E-01     0.000000E+00
    1.220000E-01     5.012400E-01     1.000000E+00
H  P
    7.270000E-01     1.0000000
H  S
    2.974000E-02     1.0000000
```

This basis set is the same as that in ref. (*44*). The additional diffuse functions are added to ccpvdz. The exponents for the diffuse functions are taken to be the same as in the aug-cc-pvdz basis sets but no diffuse polarization functions are included (O: s(0.07896), p(0.06856) and H: s(0.02974), where the number in parentheses is the exponents of the augmented orbital).

**S3.3.3 Basis set for BH**

```
B   S
     4.570000E+03      6.960000E-04     -1.390000E-04     0.000000E+00
     6.859000E+02      5.353000E-03     -1.097000E-03     0.000000E+00
     1.565000E+02      2.713400E-02     -5.444000E-03     0.000000E+00
     4.447000E+01      1.013800E-01     -2.191600E-02     0.000000E+00
     1.448000E+01      2.720550E-01     -5.975100E-02     0.000000E+00
     5.131000E+00      4.484030E-01     -1.387320E-01     0.000000E+00
     1.898000E+00      2.901230E-01     -1.314820E-01     0.000000E+00
     3.329000E-01      1.432200E-02      5.395260E-01     0.000000E+00
     1.043000E-01     -3.486000E-03      5.807740E-01     1.000000E+00
B   P
     6.001000E+00      3.548100E-02      0.000000E+00
     1.241000E+00      1.980720E-01      0.000000E+00
     3.364000E-01      5.052300E-01      0.000000E+00
     9.538000E-02      4.794990E-01      1.000000E+00
B   D
     3.430000E-01      1.0000000
B   S
     3.105000E-02      1.0000000
B   S
     9.244000E-03      1.0000000
B   P
     2.378000E-02      1.0000000
B   P
     5.129000E-03      1.0000000
B   D
     9.040000E-02      1.0000000
B   D
     2.383000E-02      1.0000000
H   S
     1.301000E+01      1.968500E-02      0.000000E+00
     1.962000E+00      1.379770E-01      0.000000E+00
     4.446000E-01      4.781480E-01      0.000000E+00
     1.220000E-01      5.012400E-01      1.000000E+00
H   P
     7.270000E-01      1.0000000
H   S
     2.970000E-02      1.0000000
H   S
     7.250000E-03      1.0000000
H   P
     1.410000E-01      1.0000000
H   P
     2.735000E-02      1.0000000
```

This basis set is the same as that in ref. (*43*). For B, two s, two p, and two d functions with exponents 0.03105 and 0.009244 for s, 0.02378 and 0.005129 for p, and 0.0904 and 0.02383 for d are added to ccpvdz. For H, two s and two p functions with exponents 0.0297 and 0.00725 for s and 0.141 and 0.02735 for p are added to ccpvdz.

**S3.3.4 Basis set for C$_2$**

```
C   S
     6.665000E+03      6.920000E-04     -1.460000E-04      0.000000E+00
     1.000000E+03      5.329000E-03     -1.154000E-03      0.000000E+00
     2.280000E+02      2.707700E-02     -5.725000E-03      0.000000E+00
     6.471000E+01      1.017180E-01     -2.331200E-02      0.000000E+00
     2.106000E+01      2.747400E-01     -6.395500E-02      0.000000E+00
     7.495000E+00      4.485640E-01     -1.499810E-01      0.000000E+00
     2.797000E+00      2.850740E-01     -1.272620E-01      0.000000E+00
     5.215000E-01      1.520400E-02      5.445290E-01      0.000000E+00
     1.596000E-01     -3.191000E-03      5.804960E-01      1.000000E+00
C   P
     9.439000E+00      3.810900E-02      0.000000E+00
     2.002000E+00      2.094800E-01      0.000000E+00
     5.456000E-01      5.085570E-01      0.000000E+00
     1.517000E-01      4.688420E-01      1.000000E+00
C   D
     5.500000E-01      1.0000000
C   S
     4.690000E-02      1.0000000
C   P
     4.041000E-02      1.0000000
```

This basis set is the same as that in ref. (*44*). The additional diffuse functions are added to ccpvdz. The exponents for the diffuse functions are taken to be the same as in the aug-cc-pvdz basis sets but no diffuse polarization functions are included (C: s(0.0469), p(0.04041), where the number in parentheses is the exponents of the augmented orbital).

## Section S4. Approximate energy conservation in SBCI1

To investigate the properties of the classical mechanical system introduced in SBCI, we examine the energy conservation in SBCI1. Since $H_{\text{SBCI1}}$ defined in Eq. 1 depends explicitly on time through $b_t^\alpha$ and $c_t^\alpha$, the energy is not exactly conserved. However, we numerically found that at discretized time steps, the following approximate energy conservation holds before and after each update:

$$\frac{b_{t-2}^\alpha}{2} y_{t-1}^{\alpha T} M y_{t-1}^\alpha + \frac{c_{t-2}^\alpha}{2} \frac{x_{t-1}^{\alpha T} H x_{t-1}^\alpha}{x_{t-1}^{\alpha T} x_{t-1}^\alpha} \approx \frac{b_{t-2}^\alpha}{2} y_t^{\alpha T} M y_t^\alpha + \frac{c_{t-2}^\alpha}{2} \frac{x_t^{\alpha T} H x_t^\alpha}{x_t^{\alpha T} x_t^\alpha} \qquad (S131)$$

Both the sides of Eq. S131 correspond to $H_{\text{SBCI1}}$ in Eq. 1, after rewriting the kinetic energy using $y_t^\alpha$ instead of $y'^\alpha_t$ and appropriately changing the subscripts of the variables. The formulation of Eq. S131 is reasonable upon the following consideration. Considering the updating rule of SBCI1 (Eqs. 10 and 11), $x_{t-1}^\alpha$ is determined by $b_{t-2}^\alpha$, and $y_{t-1}^\alpha$ is determined by $c_{t-2}^\alpha$. Therefore, it is natural to consider the SBCI1 Hamiltonian with the parameters $b_{t-2}^\alpha$ and $c_{t-2}^\alpha$ at the position $x_{t-1}^\alpha$ as follows:

$$H_{\text{SBCI1}}\{b_{t-2}^\alpha, c_{t-2}^\alpha; x_{t-1}^\alpha, y_{t-1}^\alpha\} = \frac{b_{t-2}^\alpha}{2} y_{t-1}^{\alpha T} M y_{t-1}^\alpha + \frac{c_{t-2}^\alpha}{2} \frac{x_{t-1}^{\alpha T} H x_{t-1}^\alpha}{x_{t-1}^{\alpha T} x_{t-1}^\alpha} \qquad (S132)$$

where we write the kinetic energy using $y_{t-1}^\alpha$ instead of $y'^\alpha_{t-1}$. The parameters $b_{t-2}^\alpha$ and $c_{t-2}^\alpha$ that appear explicitly in Eq. S132 can be regarded as parameters controlling the kinetic and potential energies of the dynamical system. Accordingly, when considering energy conservation in the dynamical system, these parameters should be constant. Therefore, we next consider the SBCI1 Hamiltonian with the same parameters $b_{t-2}^\alpha$ and $c_{t-2}^\alpha$, at the *updated* position $x_t^\alpha$ as follows:

$$H_{\text{SBCI1}}\{b_{t-2}^\alpha, c_{t-2}^\alpha; x_t^\alpha, y_t^\alpha\} = \frac{b_{t-2}^\alpha}{2} y_t^{\alpha T} M y_t^\alpha + \frac{c_{t-2}^\alpha}{2} \frac{x_t^{\alpha T} H x_t^\alpha}{x_t^{\alpha T} x_t^\alpha} \qquad (S133)$$

Assuming that the energy of this dynamical system is approximately conserved, we obtain Eq. S131.

We define the differences in the Rayleigh quotient and kinetic energy, $\Delta E_t^\alpha$ and $\Delta T_t^\alpha$, respectively, as

$$\Delta E_t^\alpha = \frac{x_{t-1}^{\alpha T} H x_{t-1}^\alpha}{x_{t-1}^{\alpha T} x_{t-1}^\alpha} - \frac{x_t^{\alpha T} H x_t^\alpha}{x_t^{\alpha T} x_t^\alpha} \qquad (S134)$$

$$\Delta T_t^\alpha = \frac{b_{t-2}^\alpha}{2} (y_t^{\alpha T} M y_t^\alpha - y_{t-1}^{\alpha T} M y_{t-1}^\alpha) \qquad (S135)$$

$\Delta E_t^\alpha$ is a familiar quantity used in the convergence criterion (Eqs. S30 and S61 of the Supplementary Materials for details). From Eqs. S131, S134, and S135, we obtain the following expression:

$$\Delta E_t^\alpha \approx \frac{2}{c_{t-2}^\alpha} \Delta T_t^\alpha \qquad (S136)$$

Thus, the approximate energy conservation in Eq. S131 can be verified by verifying Eq. S136. Figures S10A and S10B of the Supplementary Materials show $\Delta E_t^\alpha$ and $\frac{2}{c_{t-2}^\alpha} \Delta T_t^\alpha$ at each step $t$ for the ground state of N$_2$ ($R=$ 1.0943384670 Å). $\Delta E_t^\alpha$ and $\frac{2}{c_{t-2}^\alpha} \Delta T_t^\alpha$ are very close to each other at each step. Thus, it has been confirmed that Eq. S136 is valid, and therefore the approximate energy conservation in Eq. S131 holds. Time evolutions of parameters $b_t^\alpha$, $c_t^\alpha$ and $|x_t^\alpha|$ ($\alpha = 0$) are shown in Fig. S11 of the Supplementary Materials.

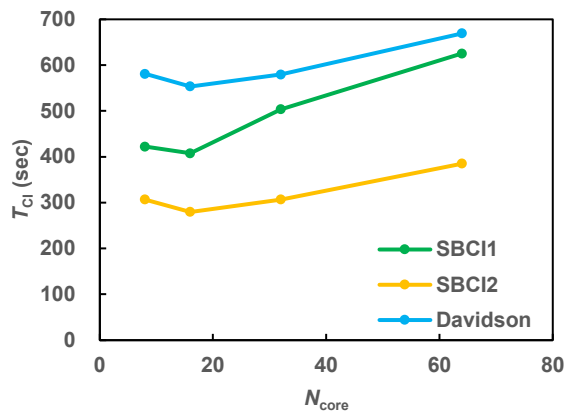

**Fig. S1. Execution time of CI, $T_{CI}$ (sec), with the different number of cores for Ne ($D_{2h}$, $A_g$, $S_z$=0, $n$=9) by SBCI1, SBCI2, and Davidson method.** Numerical values are shown in Table S1 of the Supplementary Materials.

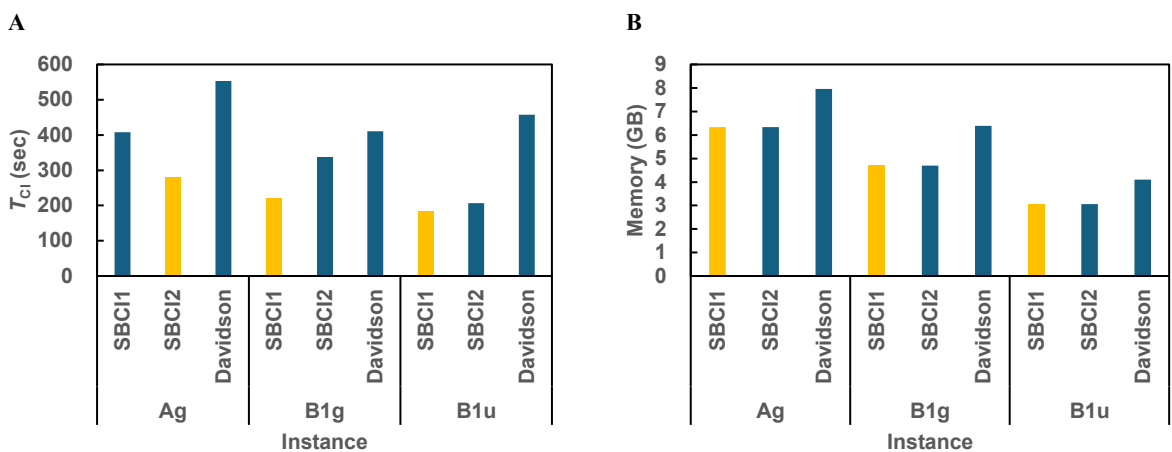

**Fig. S2. Performance of SBCI1 and SBCI2 for excited states of Ne.** (**A**) Execution time of CI, $T_{CI}$ (sec). (**B**) The amount of memory (GB) used in each instance. The values are provided in Table S5 of the Supplementary Materials.

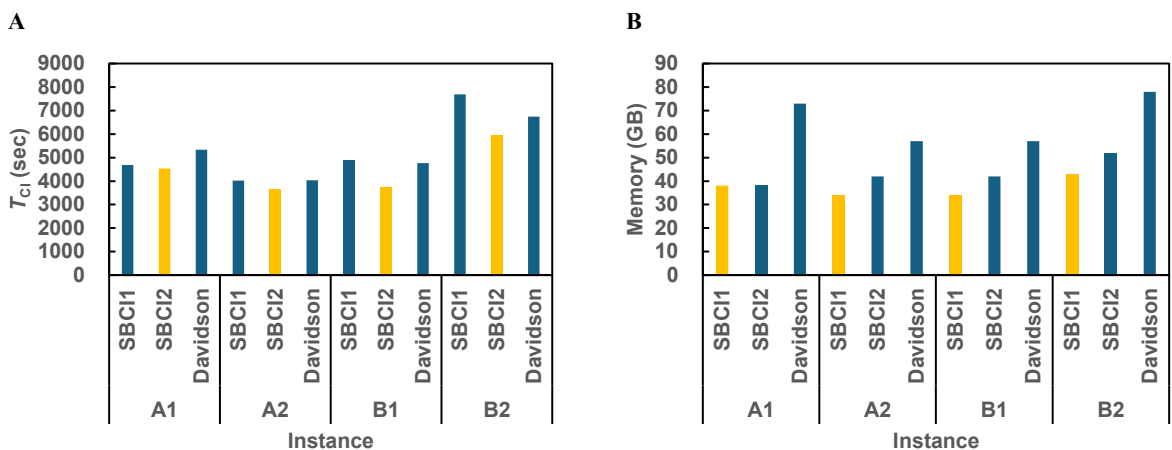

**Fig. S3. Performance of SBCI1 and SBCI2 for excited states of H$_2$O.** (**A**) Execution time of CI, $T_{CI}$ (sec). (**B**) The amount of memory (GB) used in each instance. The values are provided in Table S7 of the Supplementary Materials.

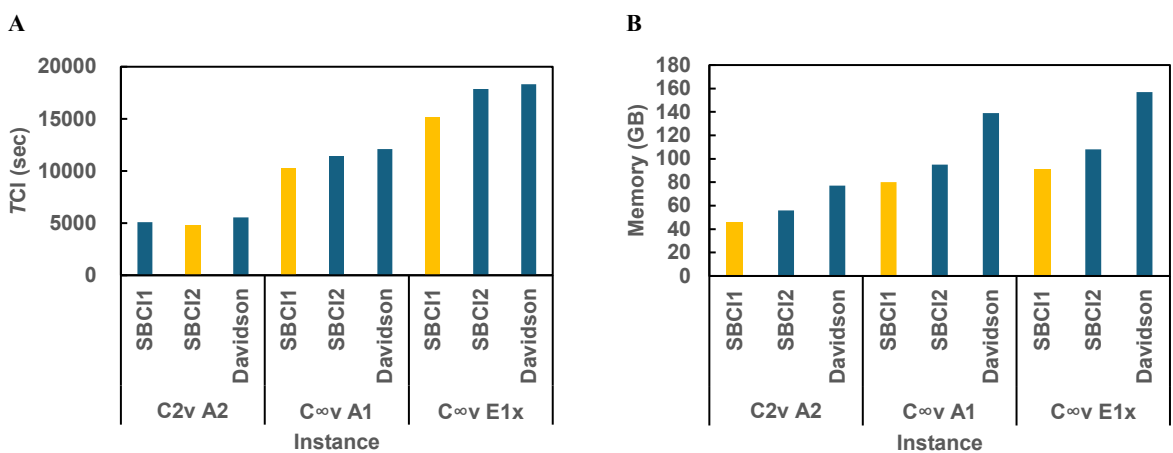

**Fig. S4. Performance of SBCI1 and SBCI2 for excited states of HF.** (**A**) Execution time of CI, $T_{CI}$ (sec). (**B**) The amount of memory (GB) used in each instance. The values are provided in Table S9 of the Supplementary Materials.

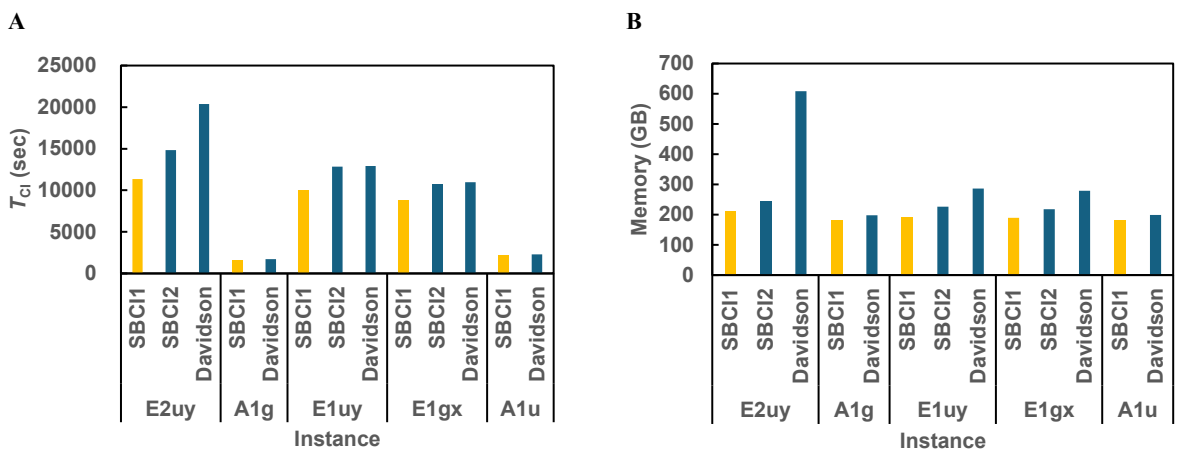

**Fig. S5. Performance of SBCI1 and SBCI2 for excited states of N$_2$.** (**A**) Execution time of CI, $T_{CI}$ (sec). (**B**) The amount of memory (GB) used in each instance. The values are provided in Table S11 of the Supplementary Materials.

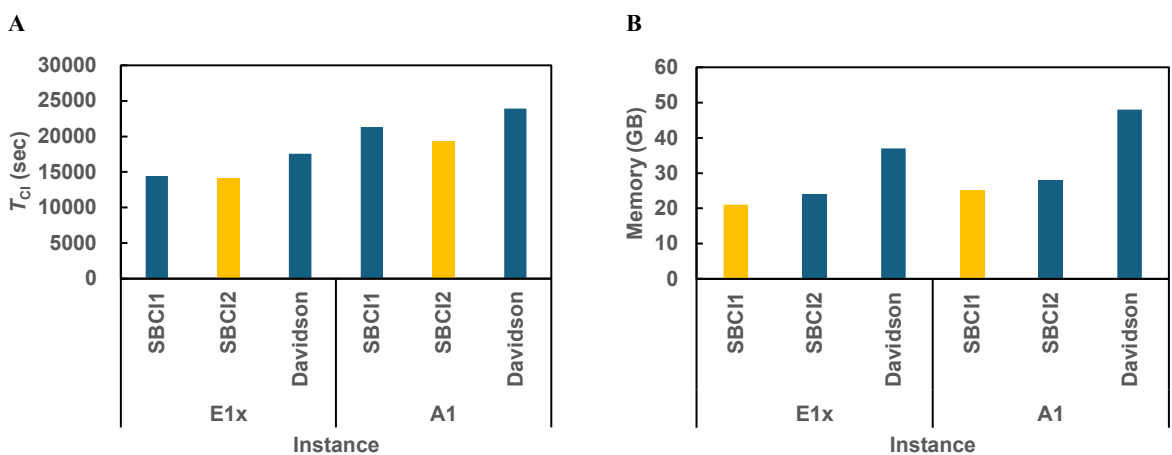

**Fig. S6. Performance of SBCI1 and SBCI2 for excited states of BH.** (**A**) Execution time of CI, $T_{CI}$ (sec). (**B**) The amount of memory (GB) used in each instance. The values are provided in Table S13 of the Supplementary Materials.

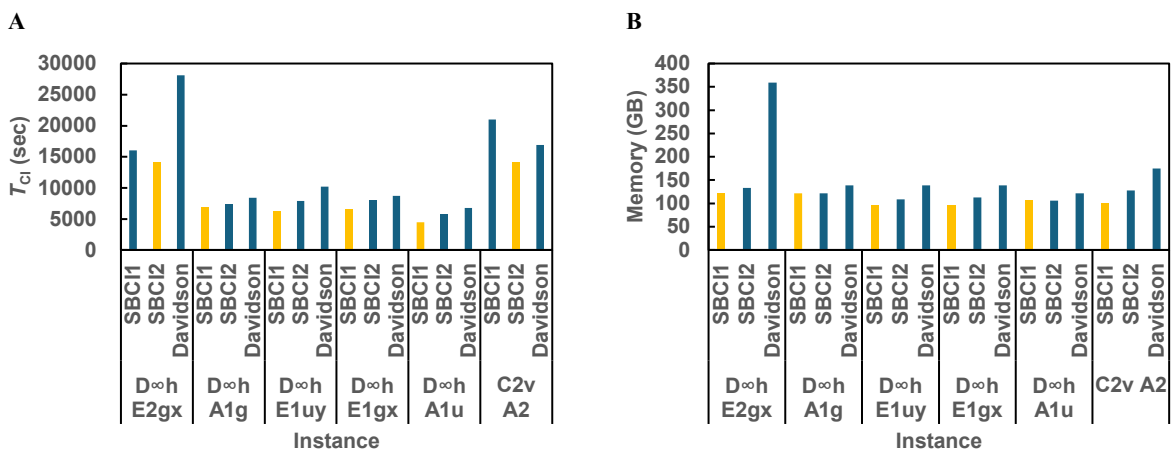

**Fig. S7. Performance of SBCI1 and SBCI2 for excited states of $C_2$.** (**A**) Execution time of CI, $T_{CI}$ (sec). (**B**) The amount of memory (GB) used in each instance. The values are provided in Table S15 of the Supplementary Materials.

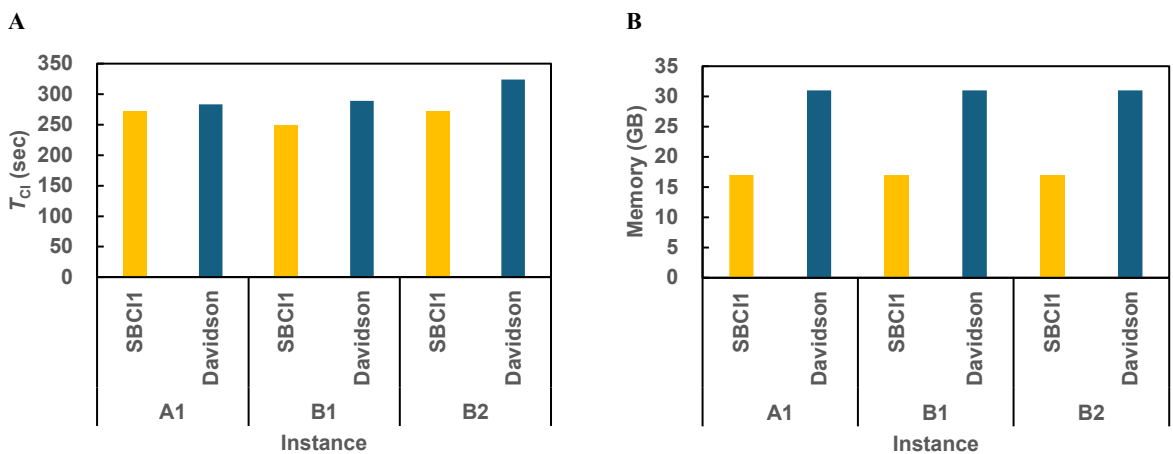

**Fig. S8. Performance of SBCI1 and SBCI2 for $H_2O^+$.** (**A**) Execution time of CI, $T_{CI}$ (sec). (**B**) The amount of memory (GB) used in each instance. The values are provided in Table S17 of the Supplementary Materials.

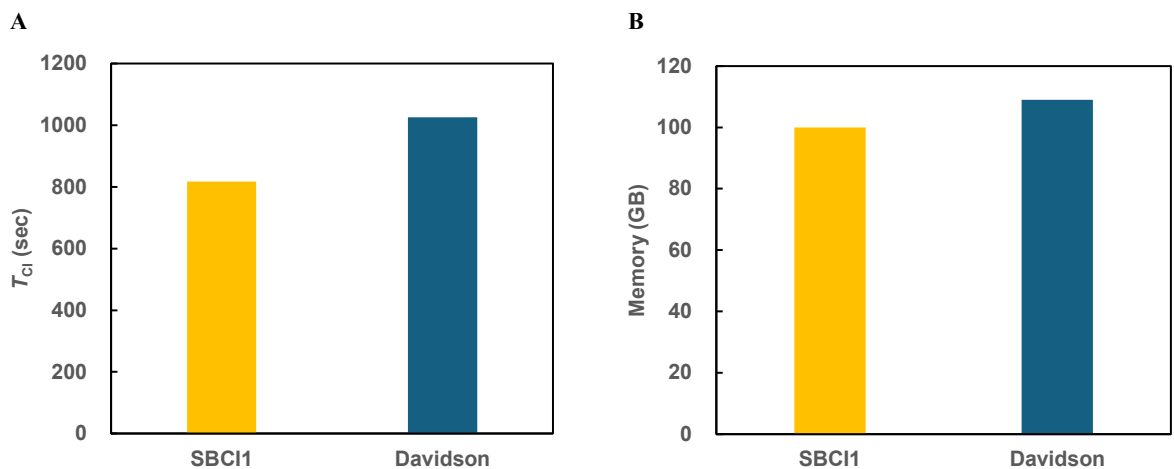

**Fig. S9. Performance of SBCI1 and SBCI2 for F$_2$.** (**A**) Execution time of CI, $T_{CI}$ (sec). (**B**) The amount of memory (GB) used in each instance. The values are provided in Table S19 of the Supplementary Materials.

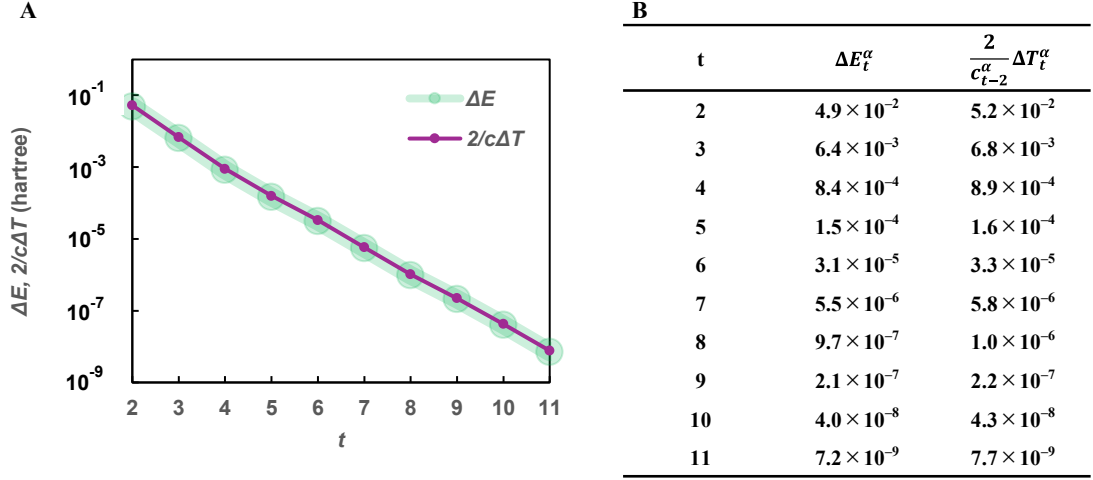

**Fig. S10. Verification of approximate energy conservation in Eq. S136 before and after the update in SBCI1.** (A) $\Delta E_t^\alpha$ and $\frac{2}{c_{t-2}^\alpha}\Delta T_t^\alpha$ in the left-hand and right-hand sides of Eq. S136, respectively, at each step $t$ for the ground state of $N_2$ at $R=$ 1.0943384670 Å. (B) Numerical values for $\Delta E_t^\alpha$ and $\frac{2}{c_{t-2}^\alpha}\Delta T_t^\alpha$.

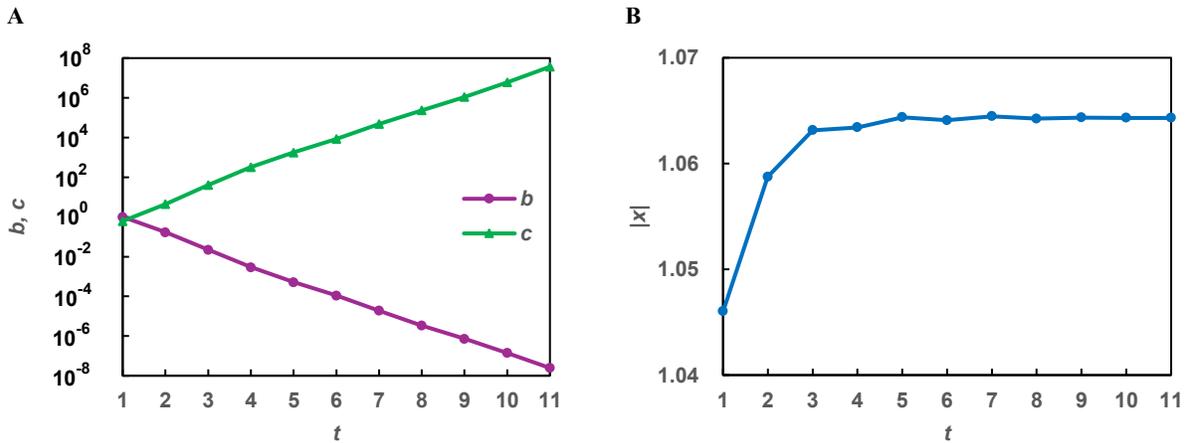

**Fig. S11. Time evolutions of parameters in Fig. S10 for the verification of approximate energy conservation in SBCI1.** (A) Parameters $b_t^\alpha$, $c_t^\alpha$ in SBCI1 at each step $t$. (B) $|x_t^\alpha|$ ($\alpha = 0$) in SBCI1 at each step $t$.

**Table S1. Execution time of CI $T_{CI}$ (sec) with the different number of cores for Ne ($D_{2h}$, $A_g$, $S_z=0$, $n=9$) by SBCI1, SBCI2, and Davidson method.**

| $N_{core}$ | SBCI1 | SBCI2 | Davidson |
|---|---|---|---|
| 8 | 422 | 307 | 581 |
| 16 | 408 | 280 | 554 |
| 32 | 504 | 307 | 579 |
| 64 | 625 | 385 | 669 |

**Table S2. Ground-state FCI total energies $E\_SBCI1$ (hartree) for various bond lengths for $N_2$ by SBCI1.** $\Delta E$(SBCI 1- Davidson) (hartree) represents the difference obtained by subtracting the energy calculated using PySCF's Davidson method from $E\_SBCI1$. $E\_ref$ (hartree) is the value of energy in ref. (*46*). Execution times of CI, $T_{CI}$ (sec), and the numbers of steps in the CI calculation ($N_{step}$) by SBCI1 and Davidson method are provided for each bond length.

| R (Å) | E_SBCI1 | ΔE(SBCI1 - Davidson) | E_ref | $T_{CI}$ SBCI1 | $T_{CI}$ Davidson | $N_{step}$ SBCI1 | $N_{step}$ Davidson |
|---|---|---|---|---|---|---|---|
| 0.793765812 | -108.6247012336 | $2.4 \times 10^{-10}$ | -108.624700 | 1466 | 1530 | 9 | 9 |
| 0.952518975 | -109.1675726468 | $2.0 \times 10^{-10}$ | -109.167573 | 1562 | 1733 | 10 | 10 |
| 1.094338467 | -109.2765275488 | $6.3 \times 10^{-10}$ | -109.276528 | 1649 | 1801 | 11 | 11 |
| 1.111272137 | -109.2781352529 | $-8.5 \times 10^{-10}$ | -109.278135 | 1766 | 1813 | 12 | 11 |
| 1.120797327 | -109.2783389661 | $-1.0 \times 10^{-9}$ | -109.278339 | 1773 | 1820 | 12 | 11 |
| 1.147256188 | -109.2765830503 | $1.0 \times 10^{-10}$ | -109.276583 | 1766 | 2029 | 12 | 13 |
| 1.173715048 | -109.2719146670 | $-1.7 \times 10^{-9}$ | -109.271915 | 1838 | 2004 | 13 | 13 |
| 1.270025300 | -109.2383971135 | $1.0 \times 10^{-9}$ | -109.238397 | 2018 | 2326 | 14 | 16 |
| 1.428778462 | -109.1603050192 | $-6.3 \times 10^{-10}$ | -109.160305 | 2376 | 2679 | 18 | 19 |
| 1.587531625 | -109.0862084246 | $9.7 \times 10^{-10}$ | -109.086211 | 3332 | 4361 | 27 | 33 |
| 1.746284787 | -109.0302893523 | $-7.2 \times 10^{-9}$ | -109.030310 | 3616 | 4775 | 30 | 37 |
| 1.905037950 | -108.9949057215 | $-2.9 \times 10^{-9}$ | -108.994810 | 4738 | 6209 | 41 | 49 |

**Table S3. Ground-state FCI total energies $E\_SBCI1$ (hartree) for various bond lengths for CN by SBCI1.** $\Delta E$(SBCI 1- Davidson) (hartree) represents the difference obtained by subtracting the energy calculated using PySCF's Davidson method from $E\_SBCI1$. $E\_ref$ (hartree) is the value of energy in ref. (48). Execution times of CI, $T_{CI}$ (sec), and the numbers of steps in the CI calculation ($N_{step}$) by SBCI1 and Davidson method are provided for each bond length.

| R (Å) | E_SBCI1 | ΔE(SBCI1 - Davidson) | E_ref | $T_{CI}$ | | $N$step | |
|---|---|---|---|---|---|---|---|
| | | | | SBCI1 | Davidson | SBCI1 | Davidson |
| 0.9 | -92.1697317616 | $-2.3 \times 10^{-11}$ | -92.169732 | 791 | 905 | 14 | 15 |
| 1 | -92.3840322443 | $1.7 \times 10^{-10}$ | -92.384032 | 857 | 958 | 15 | 16 |
| 1.0918 | -92.4693139830 | $5.4 \times 10^{-10}$ | -92.469313943 | 901 | 1007 | 16 | 17 |
| 1.1318 | -92.4856776737 | $2.2 \times 10^{-10}$ | -92.485677652 | 939 | 1063 | 17 | 18 |
| 1.1518 | -92.4904324270 | $1.8 \times 10^{-10}$ | -92.490432414 | 945 | 1065 | 17 | 18 |
| 1.1569 | -92.4913271099 | $-2.0 \times 10^{-9}$ | -92.491327096 | 1002 | 1064 | 18 | 18 |
| 1.1718 | -92.4932624230 | $-5.5 \times 10^{-10}$ | -92.493262415 | 997 | 1116 | 18 | 19 |
| 1.1769 | -92.4937049517 | $-6.3 \times 10^{-10}$ | -92.493704946 | 1004 | 1116 | 18 | 19 |
| 1.1869 | -92.4942679812 | $-8.3 \times 10^{-10}$ | -92.494267979 | 992 | 1122 | 18 | 19 |
| 1.1918 | -92.4944029632 | $-9.5 \times 10^{-10}$ | -92.494402963 | 992 | 1121 | 18 | 19 |
| 1.1919 | -92.4944047850 | $-9.5 \times 10^{-10}$ | -92.494404785 | 996 | 1115 | 18 | 19 |
| 1.1969 | -92.4944493565 | $-1.1 \times 10^{-9}$ | -92.494449358 | 992 | 1119 | 18 | 19 |
| 1.2019 | -92.4944047709 | $-1.2 \times 10^{-9}$ | -92.494404774 | 995 | 1116 | 18 | 19 |
| 1.2069 | -92.4942740208 | $-1.4 \times 10^{-9}$ | -92.494274026 | 993 | 1118 | 18 | 19 |
| 1.2118 | -92.4940650961 | $-1.6 \times 10^{-9}$ | -92.494065103 | 995 | 1118 | 18 | 19 |
| 1.2169 | -92.4937656008 | $-1.5 \times 10^{-9}$ | -92.493765608 | 1040 | 1178 | 19 | 20 |
| 1.2369 | -92.4918378206 | $-2.2 \times 10^{-9}$ | -92.491837833 | 1039 | 1176 | 19 | 20 |
| 1.2518 | -92.4896919030 | $4.1 \times 10^{-11}$ | -92.489691918 | 1037 | 1233 | 19 | 21 |
| 1.3 | -92.4793613702 | $3.6 \times 10^{-10}$ | -92.479361 | 1080 | 1292 | 20 | 22 |
| 1.4 | -92.4471465680 | $1.3 \times 10^{-11}$ | -92.447147 | 1082 | 1229 | 20 | 21 |
| 1.5 | -92.4086577177 | $-5.6 \times 10^{-9}$ | -92.408657 | 1318 | 1453 | 25 | 25 |
| 1.6 | -92.3700222418 | $-6.4 \times 10^{-9}$ | -92.370048 | 1751 | 2102 | 34 | 37 |
| 1.7577 | -92.3346161747 | $4.0 \times 10^{-9}$ | -92.316388 | 2709 | 3082 | 56 | 66 |
| 2.05065 | -92.2556694134 | $9.5 \times 10^{-9}$ | -92.255688 | 3895 | 4927 | 79 | 90 |

**Table S4. Ground-state and excited state FCI total energies $E\_SBCI2$ (hartree) for Ne by SBCI2.** $S^2$ is the expectation value of the square of the spin operator. $\Delta E$(SBCI2- Davidson) (hartree) represents the difference obtained by subtracting the energy calculated using PySCF's Davidson method from $E\_SBCI2$. $\Delta E$ex_SBCI2 (hartree) is the excitation energy by SBCI2. $\Delta E$ex_ref (hartree) is the value of excitation energy in ref. (*43*).

Ne $D_{2h}$, $A_g$, $S_z=0$, $n=9$

| State | $S^2$ | $E\_SBCI2$ | $\Delta E$(SBCI2 - Davidson) | $\Delta E$ex_SBCI2 | $\Delta E$ex_ref |
|---|---|---|---|---|---|
| 0 | 0 | -128.6851926836 | $-3.0 \times 10^{-12}$ | | |
| 1 | 2 | -128.0228348406 | $-5.0 \times 10^{-12}$ | 0.662358 | |
| 2 | 2 | -128.0172837642 | $-5.0 \times 10^{-12}$ | 0.667909 | |
| 3 | 2 | -128.0172837642 | $-3.0 \times 10^{-12}$ | 0.667909 | |
| 4 | 0 | -128.0158861535 | $-3.0 \times 10^{-12}$ | 0.669307 | 0.6693 |
| 5 | 0 | -128.0158861535 | $-2.0 \times 10^{-12}$ | 0.669307 | 0.6693 |
| 6 | 0 | -128.0058838732 | $-5.0 \times 10^{-12}$ | 0.679309 | 0.6793 |
| 7 | 2 | -127.0973005679 | $6.0 \times 10^{-12}$ | 1.587892 | |
| 8 | 0 | -127.0663107644 | $-1.5 \times 10^{-11}$ | 1.618882 | |

Ne $D_{2h}$, $B_{1g}$, $S_z=0$, $n=6$

| State | $S^2$ | $E\_SBCI2$ | $\Delta E$(SBCI2 - Davidson) | $\Delta E$ex_SBCI2 | $\Delta E$ex_ref |
|---|---|---|---|---|---|
| 0 | 2 | -128.0172837642 | $4.0 \times 10^{-12}$ | 0.667909 | |
| 1 | 0 | -128.0158861535 | 0.0 | 0.669307 | 0.6693 |
| 2 | 2 | -128.0144346312 | $9.0 \times 10^{-12}$ | 0.670758 | |
| 3 | 0 | -128.0142972807 | $2.0 \times 10^{-12}$ | 0.670895 | 0.6709 |
| 4 | 2 | -127.1238666025 | $5.2 \times 10^{-11}$ | 1.561326 | |
| 5 | 0 | -127.0001135077 | $4.6 \times 10^{-11}$ | 1.685079 | |

Ne $D_{2h}$, $B_{1u}$, $S_z=0$, $n=3$

| State | $S^2$ | $E\_SBCI2$ | $\Delta E$(SBCI2 - Davidson) | $\Delta E$ex_SBCI2 | $\Delta E$ex_ref |
|---|---|---|---|---|---|
| 0 | 2 | -128.0898604727 | $-2.0 \times 10^{-12}$ | 0.595332 | |
| 1 | 0 | -128.0825922283 | $-4.0 \times 10^{-12}$ | 0.602600 | 0.6026 |
| 2 | 6 | -127.0391849469 | $-1.3 \times 10^{-10}$ | 1.646008 | |

**Table S5. Execution time of CI, $T_{CI}$ (sec), and the amount of memory (GB) for Ne by SBCI1, SBCI2 and Davidson method.**

| Instance | $T_{CI}$ (sec) | | | Memory (GB) | | |
|---|---|---|---|---|---|---|
| | SBCI1 | SBCI2 | Davidson | SBCI1 | SBCI2 | Davidson |
| $A_g$ | 408 | 280 | 554 | 6 | 6 | 8 |
| $B_{1g}$ | 220 | 337 | 411 | 5 | 5 | 6 |
| $B_{1u}$ | 184 | 207 | 457 | 3 | 3 | 4 |

**Table S6. Ground-state and excited state FCI total energies $E\_SBCI2$ (hartree) for $H_2O$ by SBCI2.** $S^2$ is the expectation value of the square of the spin operator. $\Delta E$(SBCI2- Davidson) (hartree) represents the difference obtained by subtracting the energy calculated using PySCF's Davidson method from $E\_SBCI2$. $\Delta E$ex_SBCI2 (hartree) is the excitation energy by SBCI2. $\Delta E$ex_ref (hartree) is the value of excitation energy in ref. (*44,47*).

$H_2O$ $C_{2v}$, $A_1$, $S_z=0$ $n=4$

| State | $S^2$ | $E\_SBCI2$ | $\Delta E$(SBCI2 - Davidson) | $\Delta E$ex_SBCI2 | $\Delta E$ex_ref |
|---|---|---|---|---|---|
| 0 | 0 | -76.2582080336 | $-2.9 \times 10^{-11}$ | | |
| 1 | 2 | -75.9112617436 | $4.0 \times 10^{-9}$ | 0.346946 | 0.3470 |
| 2 | 0 | -75.8953399376 | $3.0 \times 10^{-9}$ | 0.362868 | 0.3629 |
| 3 | 2 | -75.8603984356 | $7.9 \times 10^{-10}$ | 0.397810 | 0.3978 |

$H_2O$ $C_{2v}$, $A_2$, $S_z=0$, $n=3$

| State | $S^2$ | $E\_SBCI2$ | $\Delta E$(SBCI2 - Davidson) | $\Delta E$ex_SBCI2 | $\Delta E$ex_ref |
|---|---|---|---|---|---|
| 0 | 2 | -75.9259537278 | $1.1 \times 10^{-9}$ | 0.332254 | 0.3323 |
| 1 | 0 | -75.9197268116 | $2.5 \times 10^{-10}$ | 0.338481 | 0.3385 |
| 2 | 2 | -75.8110892995 | $-7.2 \times 10^{-10}$ | 0.447119 | |

$H_2O$ $C_{2v}$, $B_1$, $S_z=0$, $n=3$

| State | $S^2$ | $E\_SBCI2$ | $\Delta E$(SBCI2 - Davidson) | $\Delta E$ex_SBCI2 | $\Delta E$ex_ref |
|---|---|---|---|---|---|
| 0 | 2 | -75.9989160776 | $2.1 \times 10^{-9}$ | 0.259292 | 0.2593 |
| 1 | 0 | -75.9845328690 | $6.1 \times 10^{-10}$ | 0.273675 | 0.2737 |
| 2 | 2 | -75.8522385571 | $-6.3 \times 10^{-8}$ | 0.405969 | 0.4060 |

$H_2O$ $C_{2v}$, $B_2$, $S_z=0$, $n=5$

| State | $S^2$ | $E\_SBCI2$ | $\Delta E$(SBCI2 - Davidson) | $\Delta E$ex_SBCI2 | $\Delta E$ex_ref |
|---|---|---|---|---|---|
| 0 | 2 | -75.8423984966 | $9.3 \times 10^{-10}$ | 0.415810 | 0.4158 |
| 1 | 0 | -75.8314939285 | $-3.3 \times 10^{-10}$ | 0.426714 | 0.4267 |
| 2 | 2 | -75.7739075922 | $1.8 \times 10^{-9}$ | 0.484300 | |
| 3 | 0 | -75.7459803093 | $-2.2 \times 10^{-10}$ | 0.512228 | |
| 4 | 2 | -75.7311314033 | $1.3 \times 10^{-9}$ | 0.527077 | |

**Table S7. Execution time of CI, $T_{CI}$ (sec), and the amount of memory (GB) for $H_2O$ by SBCI1, SBCI2 and Davidson method.**

| Instance | $T_{CI}$ (sec) | | | Memory (GB) | | |
|---|---|---|---|---|---|---|
| | SBCI1 | SBCI2 | Davidson | SBCI1 | SBCI2 | Davidson |
| $A_1$ | 4683 | 4534 | 5336 | 38 | 38 | 73 |
| $A_2$ | 4028 | 3648 | 4041 | 34 | 42 | 57 |
| $B_1$ | 4856 | 3742 | 4770 | 34 | 42 | 57 |
| $B_2$ | 7694 | 5955 | 6746 | 43 | 52 | 78 |

**Table S8. Ground-state and excited state FCI total energies $E\_SBCI2$ (hartree) for HF by SBCI2.** $S^2$ is the expectation value of the square of the spin operator. $\Delta E$(SBCI2- Davidson) (hartree) represents the difference obtained by subtracting the energy calculated using PySCF's Davidson method from $E\_SBCI2$. $\Delta E$ex_SBCI2 (eV) is the excitation energy by SBCI2. $\Delta E$ex_ref (eV) is the value of excitation energy in ref. (*47*).

| HF $C_{2v}$, $A_2$, $S_z=1$, $n=3$ | | | | | |
|---|---|---|---|---|---|
| State | $S^2$ | $E\_SBCI2$ | $\Delta E$(SBCI2 - Davidson) | $\Delta E$ex_SBCI2 (eV) | $\Delta E$ex_ref (eV) |
| 0 | 2 | -99.7155815109 | $4.6 \times 10^{-10}$ | 14.926315 | 14.926 |
| 1 | 2 | -99.7037231002 | $8.5 \times 10^{-10}$ | 15.248998 | 15.249 |
| 2 | 2 | -99.5671385015 | $5.7 \times 10^{-10}$ | 18.965648 | |
| HF $C_{\infty v}$, $A_1$, $S_z=0$, $n=4$ | | | | | |
| State | $S^2$ | $E\_SBCI2$ | $\Delta E$(SBCI2 - Davidson) | $\Delta E$ex_SBCI2 (eV) | $\Delta E$ex_ref (eV) |
| 0 | 0 | -100.2641143893 | $-1.9 \times 10^{-10}$ | | |
| 1 | 2 | -99.7665182157 | $1.4 \times 10^{-9}$ | 13.540259 | 13.540 |
| 2 | 2 | -99.7325743051 | $8.5 \times 10^{-10}$ | 14.463918 | 14.464 |
| 3 | 0 | -99.7282356264 | $1.1 \times 10^{-9}$ | 14.581979 | |
| HF $C_{\infty v}$, $E_{1x}$, $S_z=0$, $n=5$ | | | | | |
| State | $S^2$ | $E\_SBCI2$ | $\Delta E$(SBCI2 - Davidson) | $\Delta E$ex_SBCI2 (eV) | $\Delta E$ex_ref (eV) |
| 0 | 2 | -99.8949933823 | $9.6 \times 10^{-11}$ | 10.044277 | 10.044 |
| 1 | 0 | -99.8804610319 | $-7.3 \times 10^{-10}$ | 10.439722 | |
| 2 | 2 | -99.7490970637 | $1.5 \times 10^{-9}$ | 14.014312 | 14.014 |
| 3 | 0 | -99.7417641651 | $5.1 \times 10^{-10}$ | 14.213850 | |
| 4 | 2 | -99.6918743510 | $-7.8 \times 10^{-10}$ | 15.571418 | 15.571 |

**Table S9. Execution time of CI, $T_{CI}$ (sec), and the amount of memory (GB) for HF by SBCI1, SBCI2 and Davidson method.**

| Instance | $T_{CI}$ (sec) | | | Memory (GB) | | |
|---|---|---|---|---|---|---|
| | SBCI1 | SBCI2 | Davidson | SBCI1 | SBCI2 | Davidson |
| $C_{2v}$, $A_2$ | 5109 | 4813 | 5571 | 46 | 56 | 77 |
| $C_{\infty v}$, $A_1$ | 10317 | 11474 | 12118 | 80 | 95 | 139 |
| $C_{\infty v}$, $E_{1x}$ | 15176 | 17877 | 18323 | 91 | 108 | 157 |

**Table S10. Ground-state and excited state FCI total energies $E\_SBCI2$ (hartree) for $N_2$ by SBCI2.** $S^2$ is the expectation value of the square of the spin operator. $\Delta E$(SBCI2- Davidson) (hartree) represents the difference obtained by subtracting the energy calculated using PySCF's Davidson method from $E\_SBCI2$. $\Delta E$ex_SBCI2 (hartree) is the excitation energy by SBCI2. $\Delta E$ex_ref (hartree) is the value of excitation energy in ref. (*44,47*).

$N_2$ $D_{\infty h}$, $E_{2uy}$, $S_z=0$, $n=4$

| State | $S^2$ | $E\_SBCI2$ | $\Delta E$(SBCI2 - Davidson) | $\Delta E$ex_SBCI2 | $\Delta E$ex_ref |
|---|---|---|---|---|---|
| 0 | 2 | -108.9386467449 | $1.1 \times 10^{-10}$ | 0.337881 | 0.3379 |
| 1 | 2 | -108.9090538486 | $-6.4 \times 10^{-10}$ | 0.367474 | 0.3675 |
| 2 | 0 | -108.8969563845 | $-1.1 \times 10^{-9}$ | 0.379571 | 0.3796 |
| 3 | 0 | -108.8826358743 | $3.2 \times 10^{-10}$ | 0.393892 | 0.3939 |

$N_2$ $D_{\infty h}$, $A_{1g}$, $S_z=0$, $n=1$

| State | $S^2$ | $E\_SBCI2$ | $\Delta E$(SBCI2 - Davidson) | $\Delta E$ex_SBCI2 | $\Delta E$ex_ref |
|---|---|---|---|---|---|
| 0 | 0 | -109.2765276155 | $1.7 \times 10^{-9}$ | | |

$N_2$ $D_{\infty h}$, $E_{1uy}$, $S_z=0$, $n=3$

| State | $S^2$ | $E\_SBCI2$ | $\Delta E$(SBCI2 - Davidson) | $\Delta E$ex_SBCI2 | $\Delta E$ex_ref |
|---|---|---|---|---|---|
| 0 | 2 | -108.8560604342 | $1.3 \times 10^{-9}$ | 0.420467 | 0.4204 |
| 1 | 0 | -108.7764278107 | $6.9 \times 10^{-10}$ | 0.500100 | 0.5001 |
| 2 | 6 | -108.7083249895 | $1.3 \times 10^{-9}$ | 0.568203 | |

$N_2$ $D_{\infty h}$, $E_{1gx}$, $S_z=0$, $n=3$

| State | $S^2$ | $E\_SBCI2$ | $\Delta E$(SBCI2 - Davidson) | $\Delta E$ex_SBCI2 | $\Delta E$ex_ref |
|---|---|---|---|---|---|
| 0 | 2 | -108.9765327518 | $2.1 \times 10^{-9}$ | 0.299995 | 0.3000 |
| 1 | 0 | -108.9243053477 | $5.5 \times 10^{-10}$ | 0.352222 | 0.3522 |
| 2 | 6 | -108.5963410696 | $-5.3 \times 10^{-9}$ | 0.680187 | |

$N_2$ $D_{\infty h}$, $A_{1u}$, $S_z=0$, $n=1$

| State | $S^2$ | $E\_SBCI2$ | $\Delta E$(SBCI2 - Davidson) | $\Delta E$ex_SBCI2 | $\Delta E$ex_ref |
|---|---|---|---|---|---|
| 0 | 2 | -108.9863208148 | $3.7 \times 10^{-10}$ | 0.290207 | 0.2902 |

**Table S11. Execution time of CI, $T_{CI}$ (sec), and the amount of memory (GB) for $N_2$ by SBCI1, SBCI2 and Davidson method.**

| Instance | $T_{CI}$ (sec) | | | Memory (GB) | | |
|---|---|---|---|---|---|---|
| | SBCI1 | SBCI2 | Davidson | SBCI1 | SBCI2 | Davidson |
| $E_{2uy}$ | 11355 | 14843 | 20375 | 212 | 246 | 609 |
| $A_{1g}$ | 1564 | - | 1731 | 181 | - | 198 |
| $E_{1uy}$ | 9984 | 12844 | 12923 | 190 | 227 | 287 |
| $E_{1gx}$ | 8871 | 10725 | 10955 | 190 | 218 | 279 |
| $A_{1u}$ | 2226 | - | 2290 | 181 | - | 199 |

**Table S12. Ground-state and excited state FCI total energies $E\_SBCI2$ (hartree) for BH by SBCI2.** $S^2$ is the expectation value of the square of the spin operator. $\Delta E$(SBCI2- Davidson) (hartree) represents the difference obtained by subtracting the energy calculated using PySCF's Davidson method from $E\_SBCI2$. $\Delta E$ex$\_$SBCI2 (hartree) is the excitation energy by SBCI2. $\Delta E$ex$\_$ref (hartree) is the value of excitation energy in ref. (*43*).

BH $C_{\infty v}$, $E_{1x}$, $S_z=0$, $n=6$

| State | $S^2$ | $E\_SBCI2$ | $\Delta E$(SBCI2 - Davidson) | $\Delta E$ex$\_$SBCI2 | $\Delta E$ex$\_$ref |
|---|---|---|---|---|---|
| 0 | 2 | -25.1715009671 | $8.0 \times 10^{-12}$ | 0.048245 | |
| 1 | 0 | -25.1115701820 | $7.0 \times 10^{-12}$ | 0.108175 | 0.1082 |
| 2 | 2 | -24.9474437944 | $2.1 \times 10^{-11}$ | 0.272302 | |
| 3 | 0 | -24.9449962395 | $4.4 \times 10^{-11}$ | 0.274749 | 0.2744 |
| 4 | 2 | -24.9287051806 | $9.4 \times 10^{-11}$ | 0.291040 | |
| 5 | 0 | -24.9170020586 | $-4.2 \times 10^{-11}$ | 0.302744 | 0.3028 |

BH $C_{\infty v}$, $A_1$, $S_z=0$, $n=8$

| State | $S^2$ | $E\_SBCI2$ | $\Delta E$(SBCI2 - Davidson) | $\Delta E$ex$\_$SBCI2 | $\Delta E$ex$\_$ref |
|---|---|---|---|---|---|
| 0 | 0 | -25.2197455946 | 0.0 | | |
| 1 | 0 | -25.0036573047 | $-1.0 \times 10^{-12}$ | 0.216088 | 0.2161 |
| 2 | 2 | -24.9897052115 | $4.1 \times 10^{-11}$ | 0.230040 | |
| 3 | 0 | -24.9853640113 | $6.2 \times 10^{-11}$ | 0.234382 | 0.2344 |
| 4 | 0 | -24.9626416557 | $2.5 \times 10^{-11}$ | 0.257104 | 0.2571 |
| 5 | 2 | -24.9552862050 | $-1.0 \times 10^{-12}$ | 0.264459 | |
| 6 | 0 | -24.9417485969 | $4.4 \times 10^{-11}$ | 0.277997 | 0.2778 |
| 7 | 2 | -24.9397487519 | $-1.3 \times 10^{-11}$ | 0.279997 | |

**Table S13. Execution time of CI, $T_{CI}$ (sec), and the amount of memory (GB) for BH by SBCI1, SBCI2 and Davidson method.**

| Instance | $T_{CI}$ (sec) | | | Memory (GB) | | |
|---|---|---|---|---|---|---|
| | SBCI1 | SBCI2 | Davidson | SBCI1 | SBCI2 | Davidson |
| $E_{1x}$ | 14451 | 14068 | 17553 | 21 | 24 | 37 |
| $A_1$ | 21333 | 19336 | 23927 | 25 | 28 | 48 |

**Table S14. Ground-state and excited state FCI total energies $E\_SBCI2$ (hartree) for $C_2$ by SBCI2.** $S^2$ is the expectation value of the square of the spin operator. $\Delta E$(SBCI2- Davidson) (hartree) represents the difference obtained by subtracting the energy calculated using PySCF's Davidson method from $E\_SBCI2$. $\Delta E$ex_SBCI2 (hartree) is the excitation energy by SBCI2. $\Delta E$ex_ref (hartree) is the value of excitation energy in ref. (*44*).

$C_2$ $D_{\infty h}$, $E_{2gx}$, $S_z=0$, $n=5$

| State | $S^2$ | $E\_SBCI2$ | $\Delta E$(SBCI2 - Davidson) | $\Delta E$ex_SBCI2 | $\Delta E$ex_ref |
|---|---|---|---|---|---|
| 0 | 0 | -75.7302097784 | $1.5 \times 10^{-9}$ | | |
| 1 | 0 | -75.6459483736 | $2.9 \times 10^{-9}$ | 0.084261 | 0.08426 |
| 2 | 0 | -75.6362414430 | $2.2 \times 10^{-9}$ | 0.093968 | |
| 3 | 6 | -75.5331799771 | $4.6 \times 10^{-9}$ | 0.197030 | |
| 4 | 6 | -75.4851071680 | $8.5 \times 10^{-9}$ | 0.245103 | |

$C_2$ $D_{\infty h}$, $A_{1g}$, $S_z=0$, $n=3$

| State | $S^2$ | $E\_SBCI2$ | $\Delta E$(SBCI2 - Davidson) | $\Delta E$ex_SBCI2 | $\Delta E$ex_ref |
|---|---|---|---|---|---|
| 0 | 0 | -75.7302097787 | $1.4 \times 10^{-9}$ | | |
| 1 | 0 | -75.6362414455 | $1.7 \times 10^{-10}$ | 0.093968 | |
| 2 | 6 | -75.5331799802 | $-5.6 \times 10^{-9}$ | 0.197030 | |

$C_2$ $D_{\infty h}$, $E_{1uy}$, $S_z=0$, $n=3$

| State | $S^2$ | $E\_SBCI2$ | $\Delta E$(SBCI2 - Davidson) | $\Delta E$ex_SBCI2 | $\Delta E$ex_ref |
|---|---|---|---|---|---|
| 0 | 2 | -75.7190027778 | $7.0 \times 10^{-11}$ | 0.011207 | 0.01121 |
| 1 | 0 | -75.6793231527 | $9.4 \times 10^{-11}$ | 0.050887 | 0.05089 |
| 2 | 6 | -75.4414228022 | $5.3 \times 10^{-10}$ | 0.288787 | |

$C_2$ $D_{\infty h}$, $E_{1gx}$, $S_z=0$, $n=3$

| State | $S^2$ | $E\_SBCI2$ | $\Delta E$(SBCI2 - Davidson) | $\Delta E$ex_SBCI2 | $\Delta E$ex_ref |
|---|---|---|---|---|---|
| 0 | 2 | -75.6350676047 | $2.4 \times 10^{-9}$ | 0.095142 | 0.09514 |
| 1 | 0 | -75.5650618155 | $1.8 \times 10^{-9}$ | 0.165148 | 0.1651 |
| 2 | 6 | -75.5164846825 | $1.9 \times 10^{-9}$ | 0.213725 | |

$C_2$ $D_{\infty h}$, $A_{1u}$, $S_z=0$, $n=2$

| State | $S^2$ | $E\_SBCI2$ | $\Delta E$(SBCI2 - Davidson) | $\Delta E$ex_SBCI2 | $\Delta E$ex_ref |
|---|---|---|---|---|---|
| 0 | 2 | -75.6855991060 | $9.9 \times 10^{-10}$ | 0.044611 | 0.04461 |
| 1 | 0 | -75.5243305291 | $-1.1 \times 10^{-9}$ | 0.205879 | 0.2059 |

$C_2$ $C_{2v}$, $A_2$, $S_z=0$, $n=3$

| State | $S^2$ | $E\_SBCI2$ | $\Delta E$(SBCI2 - Davidson) | $\Delta E$ex_SBCI2 | $\Delta E$ex_ref |
|---|---|---|---|---|---|
| 0 | 2 | -75.6793015237 | $4.0 \times 10^{-10}$ | 0.050908 | 0.05090 |
| 1 | 6 | -75.4851071700 | $4.2 \times 10^{-10}$ | 0.245103 | |
| 2 | 6 | -75.4556652652 | $-2.5 \times 10^{-10}$ | 0.274545 | |

**Table S15. Execution time of CI, $T_{CI}$ (sec), and the amount of memory (GB) for $C_2$ by SBCI1, SBCI2 and Davidson method.**

| Instance | $T_{CI}$ (sec) | | | Memory (GB) | | |
|---|---|---|---|---|---|---|
| | SBCI1 | SBCI2 | Davidson | SBCI1 | SBCI2 | Davidson |
| $D_{\infty h}$, $E_{2gx}$ | 16059 | 14077 | 28135 | 122 | 133 | 359 |
| $D_{\infty h}$, $A_{1g}$ | 6938 | 7272 | 8432 | 122 | 122 | 139 |
| $D_{\infty h}$, $E_{1uy}$ | 6221 | 7916 | 10175 | 95 | 109 | 139 |
| $D_{\infty h}$, $E_{1gx}$ | 6476 | 7983 | 8740 | 95 | 113 | 139 |
| $D_{\infty h}$, $A_{1u}$ | 4455 | 5678 | 6751 | 106 | 106 | 122 |
| $C_{2v}$, $A_2$ | 20976 | 14052 | 16886 | 100 | 128 | 175 |

**Table S16. FCI total energies $E\_SBCI1$ (hartree) for $H_2O^+$ by SBCI1.** $S^2$ is the expectation value of the square of the spin operator. $\Delta E$(SBCI1- Davidson) (hartree) represents the difference obtained by subtracting the energy calculated using PySCF's Davidson method from $E\_SBCI1$. $E\_ref$ (hartree) is the value of energy in ref. (*45*).

$H_2O^+$ $C_{2v}$, $A_1$, $S_z$=0.5, $n$=1

| State | $S^2$ | $E\_SBCI1$ | $\Delta E$(SBCI1 - Davidson) | $E\_ref$ |
|---|---|---|---|---|
| 0 | 0.75 | -75.7329100292 | $1.0 \times 10^{-10}$ | -75.73291 |

$H_2O^+$ $C_{2v}$, $B_1$, $S_z$=0.5, $n$=1

| State | $S^2$ | $E\_SBCI1$ | $\Delta E$(SBCI1 - Davidson) | $E\_ref$ |
|---|---|---|---|---|
| 0 | 0.75 | -75.8068923642 | $1.9 \times 10^{-11}$ | -75.80689 |

$H_2O^+$ $C_{2v}$, $B_2$, $S_z$=0.5, $n$=1

| State | $S^2$ | $E\_SBCI1$ | $\Delta E$(SBCI1 - Davidson) | $E\_ref$ |
|---|---|---|---|---|
| 0 | 0.75 | -75.5582329132 | $7.3 \times 10^{-11}$ | -75.55823 |

**Table S17. Execution time of CI, $T_{CI}$ (sec), and the amount of memory (GB) for $H_2O^+$ by SBCI1 and Davidson method.**

| Instance | $T_{CI}$ (sec) | | Memory (GB) | |
|---|---|---|---|---|
| | SBCI1 | Davidson | SBCI1 | Davidson |
| $A_1$ | 272 | 284 | 17 | 31 |
| $B_1$ | 249 | 289 | 17 | 31 |
| $B_2$ | 272 | 324 | 17 | 31 |

**Table S18. Ground-state FCI total energy *E*_SBCI1 (hartree) for F₂ by SBCI1.** $S^2$ is the expectation value of the square of the spin operator. *ΔE*(SBCI1- Davidson) (hartree) represents the difference obtained by subtracting the energy calculated using PySCF's Davidson method from *E*_SBCI1. *E*_ref (hartree) is the value of energy by the original PySCF in ref. (*3*).

| State | $S^2$ | E_SBCI1 | ΔE(SBCI1 - Davidson) | E_ref |
|---|---|---|---|---|
| 0 | 0 | -198.9080786019 | $3.6 \times 10^{-10}$ | -198.9080786023 |

**Table S19. Execution time of CI, $T_{CI}$ (sec), and the amount of memory (GB) for F₂ by SBCI1 and Davidson method.**

| $T_{CI}$ (sec) | | Memory (GB) | |
|---|---|---|---|
| SBCI1 | Davidson | SBCI1 | Davidson |
| 818 | 1026 | 100 | 109 |